\documentclass[12pt]{article}
\usepackage{amsmath,bbm,amssymb,amsthm}
\usepackage{graphicx,bm,dsfont}
\usepackage{enumerate,mathtools}
\usepackage{comment,caption,multirow}
\usepackage{natbib}
\usepackage{url,subfigure} 
\usepackage{booktabs,float}
\usepackage[colorlinks,citecolor=blue,urlcolor=blue,linkcolor = blue]{hyperref}
\usepackage{algorithm,algpseudocode}

\theoremstyle{thmstyleone}%
\newtheorem{theorem}{Theorem}
\newtheorem{proposition}{Proposition}%
\theoremstyle{thmstyletwo}%
\newtheorem{example}{Example}%
\newtheorem{remark}{Remark}%
\newtheorem{corollary}{Corollary}
\newtheorem{lemma}{Lemma}
\newtheorem{definition}{Definition}

\theoremstyle{thmstylethree}%

\newcommand{\beq}{\begin{equation}}
	\newcommand{\eeq}{\end{equation}}
\newcommand{\beas}{\begin{eqnarray*}}
	\newcommand{\eeas}{\end{eqnarray*}}
\newcommand{\bea}{\begin{eqnarray}}
	\newcommand{\eea}{\end{eqnarray}}
\newcommand{\bei}{\begin{itemize}}
	\newcommand{\eei}{\end{itemize}}
\newcommand{\ben}{\begin{enumerate}}
	\newcommand{\een}{\end{enumerate}}
\newcommand{\bet}{\begin{theorem}}
	\newcommand{\eet}{\end{theorem}}
\newcommand{\bel}{\begin{lemma}}
	\newcommand{\eel}{\end{lemma}}
\newcommand{\bep}{\begin{proposition}}
	\newcommand{\eep}{\end{proposition}}
\newcommand{\bed}{\begin{definition}}
	\newcommand{\eed}{\end{definition}}
\newcommand{\bec}{\begin{corollary}}
	\newcommand{\eec}{\end{corollary}}
\newcommand{\bex}{\begin{example}}
	\newcommand{\eex}{\end{example}}

\algrenewcommand\algorithmicrequire{\textbf{Input:}}
\algrenewcommand\algorithmicensure{\textbf{Output:}}

\usepackage{amsmath,amsfonts,dsfont,mathrsfs,bm}

\def\1{\bm{1}}

\newcommand{\rd}{\mathrm{d}}
\newcommand{\ind}{\mathds{1}}

\newcommand{\eb}{\mathbf{e}}

\newcommand{\wb}{\mathbf{w}}

\newcommand{\Xb}{\mathbf{X}}

\newcommand{\cH}{\mathcal{H}}
\newcommand{\cI}{\mathcal{I}}

\newcommand{\cM}{\mathcal{M}}
\newcommand{\cN}{\mathcal{N}}

\newcommand{\EE}{\mathbb{E}}

\newcommand{\II}{\mathbb{I}}

\newcommand{\PP}{\mathbb{P}}

\newcommand{\blind}{1}

\begin{document}
	
\def\spacingset#1{\renewcommand{\baselinestretch}%
	{#1}\small\normalsize} \spacingset{1}


\if1\blind
{
	\title{\bf Harnessing The Collective Wisdom:\\ 
		Fusion Learning 
		Using Decision Sequences From Diverse Sources}
	\author{Trambak Banerjee$^1$, Bowen Gang$^2$ and Jianliang He$^3$\hspace{.5cm}\\[1ex]
		$^1$University of Kansas, $^2$Fudan University and $^3$Yale University\\
	}
	\date{}
	\maketitle
} \fi

\if0\blind
{
	\bigskip
	\bigskip
	\bigskip
	\begin{center}
		{\LARGE\bf Harnessing The Collective Wisdom: Integrative False Discovery Rate Control Using Binary Decision Sequences From Diverse Sources}
	\end{center}
} \fi
\vspace{-30pt}
	\begin{abstract}
		We introduce an Integrative Ranking and Thresholding (\texttt{IRT}) framework for fusing evidence from multiple testing procedures. The key innovation is a method that transforms binary testing decisions into compound $e-$values, enabling the combination of findings across diverse data sources or studies. We demonstrate that IRT ensures overall false discovery rate (FDR) control, provided the individual studies maintain their respective FDR levels. 
       This approach is highly flexible and is a powerful alternative for fusing inferences in meta-analysis where some studies report summary statistics while the rest reveal only the rejections under a pre-specified FDR level. Extensions to alternative Type I error control measures are explored.
	\end{abstract}
\noindent%
	
\section{Introduction}
Synthesizing the collective wisdom of crowds is related to the statistical notion of fusion learning. However, fusing inferences from diverse sources\footnote{We will use the terms `data-source' and `study' interchangeably throughout this article.} is challenging for several reasons. \textit{First}, cross-source heterogeneity and potential data-sharing complicate statistical inference, often requiring strong assumptions like study independence. 
 \textit{Second}, 
many existing meta-analytic tools require continuous summary statistics, such as $p$-values. However, it is common for some studies to only report a binary list of discoveries from an FDR-controlled procedure \citep{tang2014imputation}. Under study dependence, contemporary methods are unable to coherently integrate these mixed-evidence formats. 
\textit{Third}, disparate experimental designs and modeling techniques yield outputs that are not directly comparable, posing a significant hurdle towards their integration. \textit{Fourth}, performing such integrative analyses often requires specialized statistical expertise, limiting their broader application.
	
In this work, we propose a general and flexible framework for fusing multiple statistical testing decisions, which we call \texttt{IRT} for \texttt{I}ntegrative \texttt{R}anking and \texttt{T}hresholding. \texttt{IRT} operates under the setting where from each study a triplet is available: the study-specific vector of binary accept / reject decisions on the tested hypotheses, the FDR level of the study and the hypotheses tested by the study. Under this setting, the \texttt{IRT} framework consists of two key steps: in step (1) \texttt{IRT} utilizes the binary decisions from each study to construct nonparametric evidence indices which serve as measures of evidence against the corresponding null hypotheses, and in step (2) the evidence indices from each study are fused into a single discriminatory measure representing the overall evidence against each null hypothesis. \texttt{IRT} has several distinct advantages. \textit{First}, the \texttt{IRT} framework guarantees an overall FDR control as long as the individual studies control the FDR at their desired levels. This FDR control holds under arbitrary dependence between the fused evidence indices from step (2). See Section \ref{sec:method} for more details. \textit{Second}, \texttt{IRT} is a powerful alternative for fusing inferences in meta-analytic settings where some studies report $p-$values while the rest reveal only the rejections under a pre-specified FDR level. \cite{tang2014imputation} discuss $p-$value imputation techniques in this setting assuming that all participating studies are independent. In contrast, \texttt{IRT} synthesizes inferences setting even when the studies are dependent. Section \ref{sec:examples} presents this discussion. \textit{Third}, \texttt{IRT} is extremely simple to implement and is broadly applicable without any model assumptions. This particular aspect is especially appealing because \texttt{IRT} synthesizes inferences from diverse studies irrespective of the underlying multiple testing algorithms employed by the studies.
%

The data and R-code used in this article are available at \url{https://github.com/trambakbanerjee/IRT}.
	\section{\texttt{IRT}: integrative FDR control using binary decision sequences}
	\label{sec:method}
\subsection{Notations and problem setup}
We first introduce some notations and formally define the problem setup. We then introduce the three steps of the IRT framework: evidence construction, evidence aggregation, and FDR control.

In the sequel, let $\mathbb{I}(\cdot)$ denote the indicator function that returns 1 if the condition is met and 0 otherwise, denote $\Vert\wb\Vert_p$ as the $\ell_p$-norm of vector $\wb$, $\bm I_d$ will denote the $d\times d$ identity matrix, $\cN_d(\bm \mu,\bm\Sigma)$ will represent the $d-$dimensional Gaussian distribution with mean vector $\bm \mu$ and positive definite covariance matrix $\bm\Sigma$, a $d-$dimensional column vector with all elements equal to a real constant $a$ will be denoted by $\bm a_d$, $[d]=\{1,\ldots,d\}$ and the cardinality of a set of positive integers $\mathcal I$ will be denoted by $|\mathcal I|$.

We consider the setting of meta-analysis involving $d$ studies. For each study $j \in [d]$, a set of $m_j$ hypotheses, denoted by the index set $\mathcal{M}_j$, is tested. Let $\mathcal{M} = \bigcup_{j=1}^{d} \mathcal M_j$ denote the set of all unique hypotheses across all studies, with cardinality $|\mathcal{M}| = m$. The null hypothesis corresponding to index $i$ is denoted as $H_{0i}$.
	 Let $\mathcal{H}_{0} = \{i \in \mathcal M : H_{0i} \text{ is true}\}$ be the set of true null hypotheses and $\mathcal H_{0j}=\{H_{0i} : i \in \mathcal M_j\cap\mathcal H_0\}$ be the set of true null hypotheses tested by study $j$. 
 Denote $\theta_i=\mathbb{I}(i\notin \mathcal{H}_{0} )$ the true underlying state of hypothesis $H_{0i}$.
  For each hypothesis $i \in \mathcal{M}_j$, study $j$ makes a binary decision, $\delta_{ij} \in \{0, 1\}$, where $\delta_{ij} = 1$ signifies a rejection of $H_{0i}$. The collection of decisions for study $j$ is represented by the vector $\bm{\delta}_j =(\delta_{1j}, \cdots, \delta_{m_jj}) \in \{0,1\}^{m_j}$. Denote $\|\bm{\delta}_j\|_0=\sum_{i\in\mathcal M_j}\delta_{ij}$ the total number of rejections made by study $j$.
  
 	A selection error, or false positive, occurs if study $j$ asserts that $H_{0i}$ is false when it is in fact true. 
 	A primary goal in multiple testing is to control the False Discovery Rate \citep[FDR,][]{benjamini1995controlling}, defined as the expected proportion of false positives among all selected hypotheses. Formally,
$		\text{FDR}(\bm \delta_j) = \mathbb E\left[\text{FDP}(\bm \delta_j)\right]~\text{where FDP}(\bm \delta_j) = {\sum_{i\in \mathcal M_j}(1-\theta_i)\delta_{ij}}/{\max\{\|\bm{\delta}_j\|_0,1\}}$.
The power of a testing procedure is measured by the expected proportion of true positives detected (ETP) where, $\mbox{ETP}(\bm \delta_j)=\mathbb E[\sum_{i\in\mathcal M_j} \theta_i \delta_{ij}/\max\{\sum_{i\in\mathcal M_j} \theta_i,1\}]$.

For each study $j$, we assume the triplet $\mathcal{D}_j = \{\bm{\delta}_j, \alpha_j, \mathcal M_j\}$ is available. Here, $\alpha_j \in (0, 1)$ is the pre-specified FDR level for which the guarantee $\text{FDR}(\bm{\delta}_j) \le \alpha_j$ holds. A crucial element of our setting is that we do not always have access to the original test statistics or $p$-values that produced the decisions $\bm{\delta}_j$. Our objective is to synthesize the evidence from the collection of triplets $\{\mathcal{D}_j:j\in[d]\}$ to produce a new set of rejections for the hypotheses in $\mathcal M$, while controlling the overall FDR at a user-specified level $\alpha$.
\subsection{The  \texttt{IRT} framework}
The proposed \texttt{IRT} framework involves three steps. In Step 1, \texttt{IRT} utilizes the binary decision sequence $\bm\delta_j$ from study $j$ to construct a measure of evidence against the null hypotheses. In Step 2, this evidence is aggregated into a discriminatory measure such that for each null hypothesis $H_{0i}$, a large aggregated evidence implies stronger evidence against $H_{0i}$. In Step 3, the aggregated evidence scores are used to produce a final set of discoveries with guaranteed FDR control. In what follows, we describe each of these steps in detail. 
\\[1.5ex]
\noindent\textbf{Step 1: Evidence Construction - }To build intuition, we first consider the familiar setting where study $j$ reports decisions $\bm\delta_j$ from applying the Benjamini-Hochberg (BH) procedure \citep{benjamini1995controlling} to its corresponding $p$-values, $\bm{p}^*_j = (p^*_{ij} : i \in \mathcal{M}_j)$, at FDR level $\alpha_j$. The information available to \texttt{IRT} is the triplet $\mathcal{D}_j = \{\bm{\delta}_j, \alpha_j, \mathcal M_j\}$, which notably excludes $\bm{p}^*_j$.

Define $t_j = (\alpha_j/m_j)\|\bm{\delta}_j\|_0$, which is fully determined by the available information in $\mathcal{D}_j$.
While the true $p$-value $p^*_{ij}$ is unobserved, based on the mechanics of the BH procedure we know that $p^*_{ij} \le t_j$ for any rejected hypothesis ($\delta_{ij}=1$), and $p^*_{ij} > t_j$ for any non-rejected one ($\delta_{ij}=0$). This allows us to define a conservative $p$-value:
\begin{equation}
\label{eq:conserv_pval}
p_{ij} = t_j \cdot \mathbb{I}(\delta_{ij}=1) + 1 \cdot \mathbb{I}(\delta_{ij}=0).
\end{equation}
By construction, $p^*_{ij} \le p_{ij}$, meaning $p_{ij}$ is a valid, but conservative, $p$-value for $H_{0i}$. While these conservative $p$-values could be used in traditional meta-analysis, an alternative and increasingly popular approach is to transform evidence into \emph{e-values} \citep{vovk2021values}. An $e-$value is a non-negative random variable $e$\footnote{We will use the notation `$e$' to denote both the random variable and its realized value.} with the property that $\mathbb{E}[e] \le 1$ under the null hypothesis; a large $e$ indicates strong evidence against the null. A key advantage of $e-$values over $p-$values is their ease of aggregation. In fact, \cite{vovk2022admissible} show that admissible methods for combining $p$-values under arbitrary dependence essentially operate by first converting $p$-values into $e-$values and then averaging them. In the context of our BH example, if the original $p$-values $\{p^*_{ij}\}$ are independent, it can be shown that transforming our conservative $p$-values via
\begin{equation} \label{eq:eij_1}
	e_{ij} =
	\begin{cases}
	\pi_j	p_{ij}^{-1} & \text{if } \delta_{ij} = 1 \\
		0           & \text{if } \delta_{ij} = 0
	\end{cases},
\end{equation}
where $\pi_j=|\mathcal{H}_{0j}|/m_j$, yields a valid $e-$value for each hypothesis $H_{0i}$ (see Lemma \ref{lem4} in Supplement \ref{sec:irt_star}).

The central insight of our work is that this principle extends far beyond the BH procedure and does not require the FDR procedure to use $p$-values at all. To generalize this idea to decisions from \emph{any} FDR-controlling method (such as knockoffs \citep{barber2015controlling} or covariate-powered methods \citep{ignatiadis2021covariate}) using arbitrarily dependent test statistics, we shift our perspective from individual $e$-values to \emph{compound $e-$values} \citep{wang2022false,ren2024derandomised,ignatiadis2024compound}.

\begin{definition}[Compound $e-$values] \label{def:ge}
	Let $\bm{e} =\{e_1, \dots, e_m\}$ be a collection of random variables associated with the hypotheses $H_{01}, \dots, H_{0m}$. 
    We say $\bm{e}$ is a set of compound $e-$values if $\sum_{i \in \mathcal{H}_0}\mathbb{E}[e_i] \le |\mathcal H_0|$.
\end{definition}

With this concept, we now define the general evidence construction for \texttt{IRT}, which operates directly on the triplet $\mathcal{D}_j$:
\begin{equation} \label{eq:eij}
	e_{ij} = w_j\frac{\delta_{ij}}{\max(\|\bm{\delta}_j\|_0, 1)},\text{ where the evidence weight }w_j = \frac{m_j}{\alpha_j},~\forall i \in \mathcal M_j.
\end{equation}
The term $\delta_{ij} / \max(\|\bm{\delta}_j\|_0,1)$ distributes the ``evidence weight'' of study $j$ evenly across its rejected hypotheses. The weight $w_j$ assigns greater importance to rejections from studies that are larger (more hypotheses $m_j$) or more conservative (smaller FDR level $\alpha_j$). This general construction is the cornerstone of our framework, leading to our main theoretical result for this step.
\begin{theorem} \label{thm1}
	Suppose study $j$ controls FDR at level $\alpha_j$. Then the collection of evidence indices $\bm{e}_j = \{e_{ij}\}_{i \in \mathcal M_j}$ from Equation~\eqref{eq:eij} is a set of compound $e-$values associated with the null hypotheses in $\mathcal H_{0j}$.
\end{theorem}
In Section \ref{sec:evid_motivation} of the supplement, we show that these evidence indices also naturally arise as building blocks of several popular aggregation and derandomization procedures, such as those of \cite{ren2024derandomised} and \cite{li2023values}.  Section \ref{sec:alternative_typeI} explores how compound $e-$values can be derived from decisions that control alternative notions of Type I error.
\\[1.5ex]	
\noindent\textbf{Step 2: Evidence Aggregation - }
Given the compound $e-$values from each study, 
\texttt{IRT} aggregates the evidence indices $\eb_j$ across the studies as follows:
\begin{equation}\label{eq:fede}
				{e}_i^\texttt{agg}=\dfrac{1}{d}\sum_{j=1}^{d}\Bigl\{e_{ij}~\mathbb I(i\in\mathcal M_j)+\mathbb I(i\notin \mathcal M_j)\Bigr\},\quad\forall i\in\cM.
		\end{equation}
When each study tests all the $m$ hypotheses and $m_j=m$, then $e_i^{\tt agg}$ is the arithmetic mean of the $d$ evidence indices corresponding to hypothesis $i$. However, when $m_j$ are different, the aggregation scheme in Equation \eqref{eq:fede} sets $e_{ij}=1$ whenever $i\notin\mathcal M_j$, which is a valid $e-$value\footnote{Note that in our framework, we treat the $e-$values for hypotheses in $\mathcal M\setminus \mathcal M_j$ as missing completely at random \citep{rubin1976inference}. Here `` $\setminus$ " is the usual set difference operator.}. This aggregation preserves the compound $e-$value property as the next Theorem shows.
\begin{theorem}\label{thm2}
Suppose that each study $j$ controls FDR at level $\alpha_j$. Then, $\bm e^\texttt{agg}=\{e_i^{\texttt{agg}}:i\in\mathcal M\}$ is a set of compound $e-$values associated with $\mathcal H_0$.
\end{theorem}%
The intuition for this choice is that for standard $e-$values, simple averaging is known to dominate any other symmetric aggregation function under arbitrary dependence \citep{vovk2021values, wang2024only}. While we are working with compound $e-$values of a specific form, this provides strong heuristic justification for our approach. Whether simple averaging is formally admissible in this specific setting remains an interesting open question.
\\[1.5ex]	
\noindent\textbf{Step 3: FDR control - }Once the aggregated evidence indices $\bm{e}^{\,\texttt{agg}}$ are constructed, there are two primary paths to obtain a final set of rejections, depending on the available data. The most direct approach is to apply the e-BH procedure \citep{wang2022false} to the set of aggregated compound $e-$values, $\bm{e}^{\,\texttt{agg}}$. Specifically, denote $e_{(1)}\ge\ldots\ge e_{(m)}$ as the ordered $e-$values from largest to smallest. The rejection rule of e-BH is given by 
$\delta_i=\mathbb I\{e_{i}\ge {m}/(\alpha k_\alpha)\}$ for all $i\in\cM$, where the threshold is chosen as
$k_\alpha= \max\{i\in\mathcal M:e_{(i)}\geq {m}/({i\alpha})\}$ with the convention that $\max{(\emptyset)}=0$.
This method guarantees FDR control at the desired level $\alpha$ under arbitrary dependence structures among the $e-$values.
However, a key limitation of this approach is that it cannot make rejections if the target FDR level $\alpha$ is more stringent than that of any individual study, i.e., if $\alpha < \min_{j\in[d]}\alpha_j$. This may seem counterintuitive for meta-analysis, but it is not a flaw of our method. Rather, it is an intrinsic property of any procedure that aggregates evidence derived solely from binary decisions. Without additional information or assumptions, one cannot generate evidence stronger than the strongest input. A more detailed discussion of this property is provided in Section \ref{sec:alpha_choice} of the supplement.

Alternatively, \texttt{IRT} provides a powerful building block for integrative analyses when heterogeneous data types are available. Consider a common meta-analysis setting where, in addition to the $d$ studies providing binary decisions, we have access to a separate, independent study that reports a full set of $p$-values, $\bm{p}=\{p_1,\ldots, p_m\}$. We can fuse this information with our aggregated compound $e-$values, $\bm{e}^{\texttt{agg}}$, using the \emph{ep-BH procedure} \citep{ignatiadis2024compound}. This method treats the $e-$values as unnormalized, data-driven weights for the $p$-values. Specifically, one first computes a set of re-weighted $p$-values, $p'_i = \min(p_i / e_i^{\texttt{agg}},1)$, and then applies the standard BH procedure to this new set $\{p'_i:i\in\mathcal M\}$. The primary benefit of this hybrid approach is its potential for greater statistical power compared to analyzing each data source separately. That is, the ep-BH procedure can yield more discoveries than either: (1) applying the standard BH procedure to the independent $p$-values $\bm p$ alone, or (2) applying the e-BH procedure to the aggregated $e-$values $\bm e^{\texttt{agg}}$ alone. Section \ref{sec:examples} illustrates this with numerical examples.
\section{Numerical illustrations}
\label{sec:examples}
We illustrate the utility of IRT for meta-analysis under the setting where $d_1$ studies report $p-$values while $d_2$ studies report their binary decision sequences. \cite{tang2014imputation} discuss $p-$value imputation techniques for meta-analysis under this setting but assume that the $d=d_1+d_2$ studies are independent. In the next two examples, we demonstrate that when the studies are dependent, in a sense that is discussed subsequently, \texttt{IRT} provides a powerful strategy for pooling inferences in this scenario.

Suppose, without loss of generality, the first $d_1$ studies report their raw $p-$values for each of the $m$ hypotheses while the remaining $d_2=d-d_1$ studies report the triplet $\mathcal D_j, j=d_1+1,\ldots,d$, where the corresponding decision sequences $\bm \delta_j$ are obtained from the BH procedure with FDR control level $\alpha_j=0.01$. We set $m=1000$ and consider testing $H_{0i}:\mu_i=0~vs~H_{1i}:\mu_i\ne 0$, where $\mu_i\stackrel{\rm i.i.d.}{\sim}0.95~\delta_{\{0\}}+0.025~\cN(3,1)+0.025~\cN(-3,1)$ and $\delta_{\{a\}}$ denotes a point mass at $a$. For each hypothesis $i$, the test statistics $\bm X_i=(X_{ij}:j\in[d])\stackrel{ind.}{\sim} \cN_d(\mu_i\bm 1_d,\bm\Sigma)$ where $\bm \Sigma=\begin{pmatrix}
\bm \Sigma_{d_1} & \bm 0_{d_1\times d_2}\\
\bm 0_{d_2\times d_1} & \bm \Sigma_{d_2}
\end{pmatrix}$ so that the $d_1$ and $d_2$ studies are independent of each other. We set $\bm\Sigma_{d_k}=\rho_k\bm 1_{d_k}\bm 1_{d_k}^T+(1-\rho_k)\bm I_{d_k}$ for $k\in\{1,2\}$ and $\rho\in (0,1)$. The raw $p$-values are computed using the standard two-sided $Z-$test formula, $p_{ij} = 2\Phi(-|X_{ij}|)$, where $\Phi$ is the distribution function for standard normal. The following four methods are evaluated for integrative inference at FDR level $\alpha$: (i) \textbf{Cauchy $d_1$}, a baseline which derives the pooled $p-$values from the first $d_1$ studies using the Cauchy combination test statistic \citep{liu2020cauchy} followed by a BH correction; (ii) \textbf{Cauchy}, an idealized benchmark which is similar to \textbf{Cauchy $d_1$} but derives the pooled $p-$values from all $d$ studies, (iii) \textbf{Cauchy + Imputed}, that first imputes the $p-$values for the $d_2$ studies using Equation \eqref{eq:conserv_pval}, then pools all $p-$values using the Cauchy combination test statistic and finally applies the BH correction,  (iv) \textbf{Cauchy + IRT}, which is the hybrid approach. We first use the Cauchy combination test to produce a single pooled $p$-value vector, $\bm p$, from the $d_1$ studies. We then apply \texttt{IRT} to the $d_2$ studies to generate aggregated $e-$values, $\bm e^{\texttt{agg}}$. Finally, we apply the ep-BH procedure \citep{ignatiadis2024compound} to the pairs $\{\bm p, \bm e ^{\texttt{agg}}\}$. This procedure controls FDR because the block diagonal structure of $\bm\Sigma$ ensures $\bm p$ and $\bm e ^{\texttt{agg}}$ are independent. 
The performances of these four methods are compared with the quality of inference obtained from a \textbf{single study} that applies the BH procedure on the $p-$values from the first study for FDR control at level $\alpha_1$.
\begin{example}
\label{ex:1}
We fix $d_1=2,~d_2=3,~\rho_k=\rho$ and vary $\rho\in\{0,0.1,0.3,0.5,0.7,0.9\}$. Across $2000$ Monte-Carlo (MC) repetitions of this data generating scheme, Figure \ref{fig:meta_blocdep_1} reports the average
FDP and ETP of the five different methods for integrative inference at the FDR level $0.5\%$. All methods for integrative inference control FDR at $0.5\%$ while the single study controls it at its designated FDR level $1\%$. Unsurprisingly, \texttt{Cauchy}, overall, has the highest power across almost all values of $\rho$ as it uses the $p-$value information from all $d$ studies for integrative inference. In contrast, inferences that rely on a conservative $p-$value imputation method have the least power.  This is followed by \texttt{Cauchy $d_1$} which fuses inference across the first $d_1$ studies. When $\rho$ is high, pooled inferences from \texttt{Cauchy $d_1$} are less powerful than those from a single study. This is expected since \texttt{Cauchy $d_1$} controls FDR at a more stringent level and pooling inferences across highly dependent studies may not lead to substantially more true positives than what can be learned from a single study. The key comparison is between \texttt{Cauchy $d_1$} and \texttt{Cauchy + IRT}. Across all levels of correlation, \texttt{Cauchy + IRT} is uniformly more powerful. 
This result directly demonstrates the value of the information contained in the binary decisions. By transforming them into compound $e-$values, \texttt{IRT} allows us to extract meaningful evidence and achieve greater statistical power than an analysis that discards this information.
\begin{figure}[!t]
\centering
\includegraphics[width=0.72\linewidth]{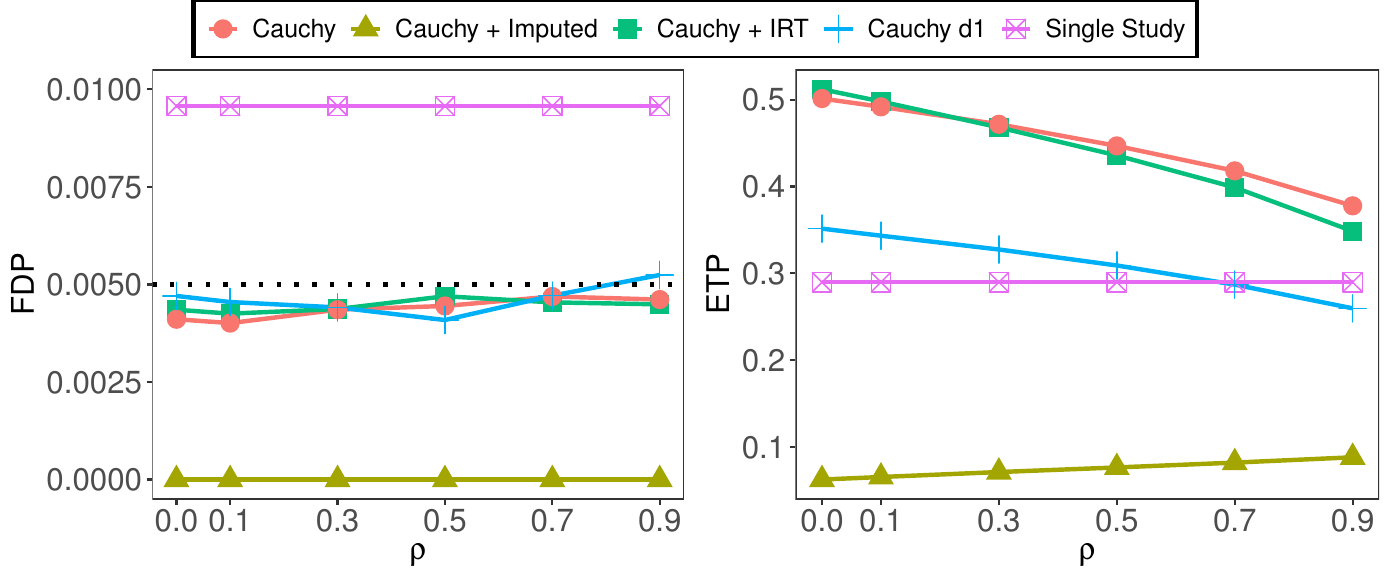}
\caption{FDP and ETP comparison for Example \ref{ex:1}.}
\label{fig:meta_blocdep_1}
\end{figure}
\end{example}
\begin{example}
\label{ex:2}
We now examine a setting with asymmetric dependence, where the $p$-value studies are correlated (with correlation $\rho_1$) while the binary-decision studies are independent ($\rho_2 = 0$). We continue to take $d_1=2,~d_2=3$ but set $\alpha=\alpha_j=0.01$ for $j\in[d]$.  Figure \ref{fig:meta_blocdep_2} reports the average
FDP and ETP for various methods across 2000 MC repetitions. While all methods control FDR at $1\%$, we find that across all values of $\rho_1$, \texttt{Cauchy + IRT} is more powerful than \texttt{Cauchy + Imputation}, \texttt{Cauchy $d_1$} and the \texttt{Single Study}.
\begin{figure}[!h]
\centering
\includegraphics[width=0.71\linewidth]{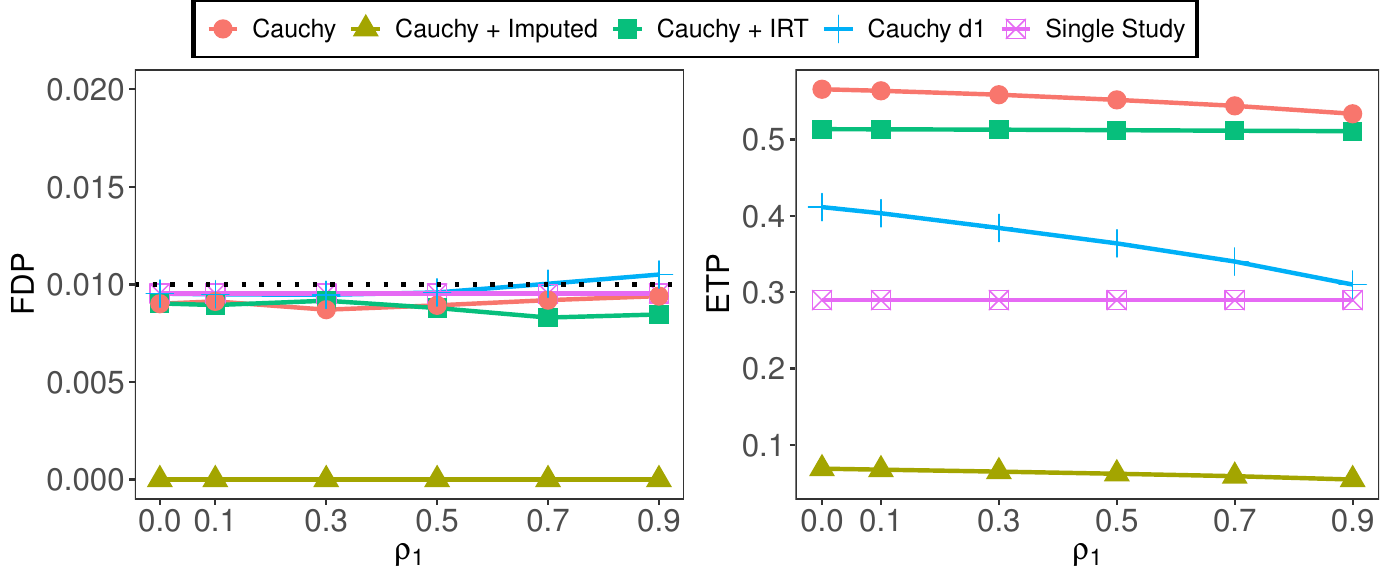}
\caption{FDP and ETP comparison for Example \ref{ex:2}.}
\label{fig:meta_blocdep_2}
\end{figure}
\end{example}
Additional numerical experiments and a real data analysis illustrating the empirical performance of \texttt{IRT} are, respectively, presented in sections \ref{sec:sims} and \ref{sec:realdata} of the Supplement.
\section{Concluding remarks}
\label{sec:summary}
\texttt{IRT} is a general framework for fusion learning in multiple testing that operates on the binary decision sequences available from diverse studies and conducts integrative inference on the common parameter of interest.  For meta-analysis involving dependent studies, \texttt{IRT} provides a powerful alternative for fusing inferences when some studies report $p-$values while the rest reveal only the rejections under a pre-specified FDR level. 
Section \ref{sec:irt_star} of the supplement proposes \texttt{IRT$^*$} and \texttt{IRT H} which are powerful alternatives to \texttt{IRT} under additional assumptions on the data generating process for each study.
		
While the focus of this article is on testing the intersection null for meta-analysis, a natural extension of our framework lies in multiple testing of partial conjunction (PC) hypotheses (see \cite{benjamini2008screening,10.1214/21-AOS2139,bogomolov2023testing} for an incomplete list of references). Here the goal is to test if at least $u\ge 1$ out of the $d$ studies reject the null hypothesis $H_{0i},~i=1,\ldots,m$, i.e., to test $H_{0i}^{u/d}:\text{fewer than $u$ out of $d$ studies are non-null}$. Given the triplet $\mathcal D_j$ from each study, a key challenge in this setting is to construct a powerful aggregation scheme such that the aggregated evidence indices provide an effective ranking of the $m$ composite PC null hypotheses. 
On a related note, the current \texttt{IRT} framework does not handle settings like \cite{zollinger2015meta} where some studies may only reveal the ranks of the top differentially expressed genes. Extending \texttt{IRT} to fuse inferences from such mixed data types across dependent studies is a promising direction for future research.

\bibliographystyle{chicago}
\bibliography{refs}
\newpage
\section*{Supplementary material}
\label{SM}
This supplement is organized as follows: Section \ref{sec:alpha_choice} discusses the impact of target FDR levels on the power of \texttt{IRT}. In Section \ref{sec:irt_star} we present \texttt{IRT$^*$} that relies on an alternative scheme for evidence aggregation. The proofs of all theoretical results in the paper are presented in Section \ref{app:proofs}. In Section \ref{sec:discuss} we present additional insights about the \texttt{IRT} framework. In particular, we show that (i) the evidence indices in Equation \eqref{eq:eij} are connected to some existing aggregation and derandomization procedures (Section \ref{sec:evid_motivation}), and (ii) prove that \texttt{IRT} guarantees asymptotic FDR control if some studies control their FDR asymptotically (Section \ref{sec:agent_fdr}). 
Section \ref{sec:alternative_typeI} extends \texttt{IRT} to alternative forms of Type I error control. Additional numerical studies and a real data application are presented in sections \ref{sec:sims} and \ref{sec:realdata}, respectively. 

\vspace*{-10pt}
\appendix
\section{Impact of $\alpha$ on the power of \texttt{IRT}}
\label{sec:alpha_choice}
When all $d$ studies report the triplets $\{\mathcal D_j:j\in[d]\}$, the choice of $\alpha$ bears important consideration as far as the power of IRT is concerned, where we refer to IRT as the procedure that applies the e-BH method directly to the aggregated e-values, $\pmb{e}^{\text{agg}}$. For instance, with a relatively smaller value of $\alpha$, \texttt{IRT} may fail to recover discoveries identified by studies with a smaller weight $w_j$. In fact, when inferences are pooled with the goal of achieving higher reliability then often $\alpha<\min_{j\in[d]}\alpha_j$, and in such settings \texttt{IRT} may exhibit no power. In this section we take a simple example to discuss the impact that $\alpha$ has on the power of \texttt{IRT}. Thereafter, in Section \ref{sec:irt_star} we present \texttt{IRT$^*$}, which relies on an alternative evidence aggregation scheme and is more powerful than \texttt{IRT} when inferences are synthesized for higher reliability.
			
\begin{figure}[!h]
				\centering
				\includegraphics[width=0.8\linewidth]{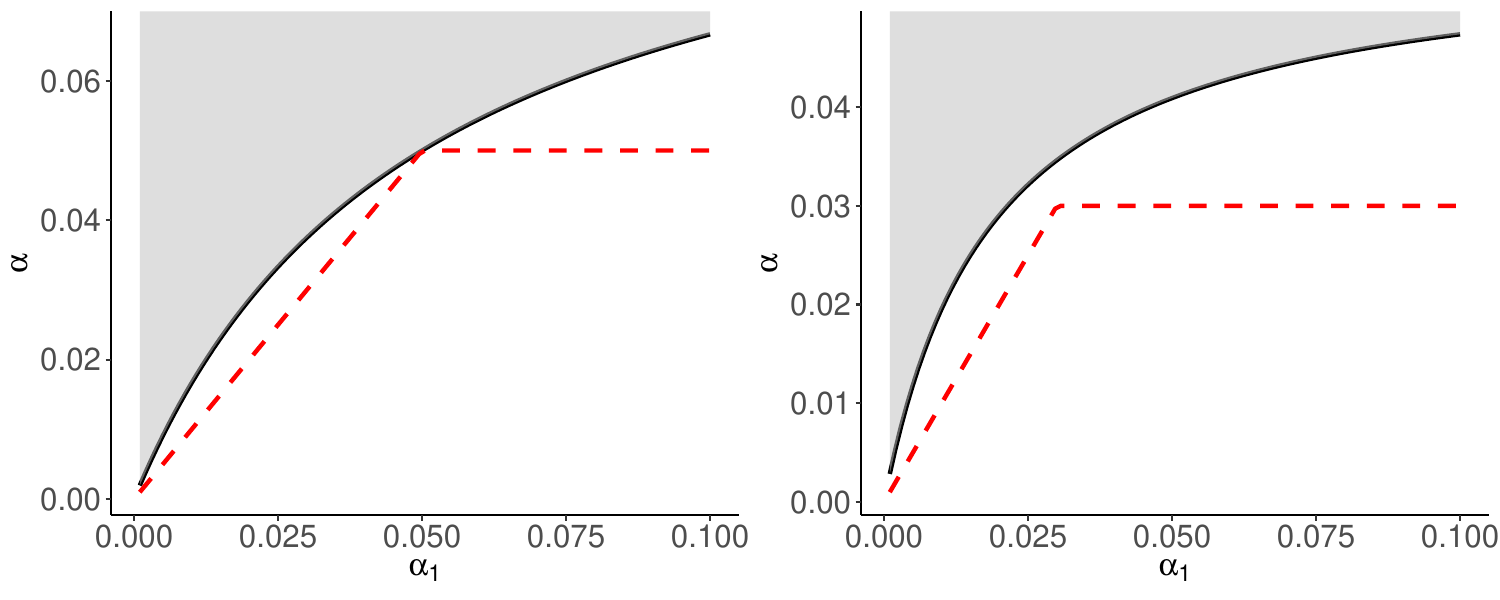}
				\caption{Pooling inferences using \texttt{IRT}. Left: Here $d=2$ studies are testing the same set of $m=5$ null hypotheses at levels $\alpha_1$ and $\alpha_2$, respectively. Both studies reject only the $i^{\tt th}$ null hypothesis. The gray shaded region depicts the overall FDR level $\alpha$ required for the e-BH procedure to reject $H_{0i}$ when $\alpha_1$ varies over $0.001$ to $0.1$ and $\alpha_2$ is fixed at $0.05$. Here the red dotted line represents $\min(\alpha_1,\alpha_2)$. Right: Same setting with $d=3,~\alpha_2=0.05$ and $\alpha_3=0.03$.}
				\label{fig:alpha_plot}
			\end{figure}

Suppose there are $d=2$ studies, each testing the same set of $m=5$ null hypotheses at levels $\alpha_1$ and $\alpha_2$, respectively. Consider a simple setting where the $d$ studies reject only the $i$-th null hypothesis. So $\delta_{ij}=1$ and $\|\bm{\delta}_j\|_0=1$ for all $j\in[d]$. Suppose \texttt{IRT} is used to pool inferences from these studies. The gray shaded region in the left panel of Figure \ref{fig:alpha_plot} depicts the overall FDR level $\alpha$ required for the e-BH procedure to reject $H_{0i}$ when $\alpha_1$ varies over $0.001$ to $0.1$ and $\alpha_2$ is fixed at $0.05$. Here the red dotted line represents $\min(\alpha_1,\alpha_2)$. Most notably, this plot reveals that one must have $\alpha>\min(\alpha_1,\alpha_2)$ to reject $H_{0i}$ unless $\alpha_1=\alpha_2=0.05$, in which case $\alpha$ must at least be $0.05$ for the e-BH procedure to reject $H_{0i}$. The right panel considers the same setting but with $d=3,~\alpha_2=0.05,~\alpha_3=0.03$ and paints a similar picture.

			The calculations for Figure \ref{fig:alpha_plot} readily follow from the \texttt{IRT} procedure. For instance, in the case of the left panel, \texttt{IRT} rejects $H_{0i}$ at FDR level $\alpha$ if $\alpha\ge 2\alpha_1\alpha_2/(\alpha_1+\alpha_2)$. We have $2\alpha_1\alpha_2/(\alpha_1+\alpha_2)\ge \min(\alpha_1,\alpha_2)$ where equality holds when $\alpha_1=\alpha_2$.  The results from the left panel may seem counter-intuitive, as the decisions of the second study do not necessarily enhance the evidence against $H_{0i}$. However, this in fact reflects a fundamental constraint inherent to the nature of the problem. If the two studies use identical data and methods, the collective evidence may not be stronger than the individual evidence, since the latter study does not provide fresh information.
            No fusion learning method, including IRT, could justifiably claim a rejection at a lower FDR level than $\min\{\alpha_1,\alpha_2\}$. The IRT framework is designed with the explicit goal of guaranteeing FDR control under minimal assumptions. This principle requires the framework to be conservative enough to remain valid even in worst-case scenarios of inter-study dependence. 
            

\section{\texttt{IRT$^*$}: more powerful evidence aggregation via multiplication}
\label{sec:irt_star}
If inferences are pooled with the goal of achieving higher reliability then an important implicit assumption is that, informally, the studies are ``different'' in some sense. In this section we make this idea precise and present \texttt{IRT$^*$}, which relies on an alternative evidence aggregation scheme.
We will need the following definitions.
\begin{definition}[Partial exchangeability]
				\label{def:partial_exchang}
				Let $\{X_i\}_{i\in\cM}$ be a set of random variables and $\cI_0$ a subset of $\cM$.
				We say $\Xb=\{X_i\}_{i\in\cM}$ is partially exchangeable on $\cI_0$  if  
				$f(\Xb)=f(\Psi_{i,i'}\{\Xb\})$ for all $i,i'\in\cI_0$, where $\Psi_{i,j}$ is the permutation that swaps the $i$-th and the $j$-th positions, and $f$ is the joint density function of $\Xb$.  
			\end{definition}
			\begin{definition}[Symmetric decision rule, \cite{copas1974symmetric}]
				\label{def:symmetry}
				A decision rule $\bm{\delta}$ is \textit{symmetric} if
				$\bm{\delta}(\Psi\{{\Xb}\})=\Psi\{\bm{\delta}(\Xb)\}$
				for all permutation operators $\Psi$.
			\end{definition}
			The notion of partial exchangeablilty on the set of nulls is commonly used in the conformal inference literature \citep{bates2023testing,liang2024integrative} while symmetric decision rules arise naturally in conventional settings where all hypotheses undergo simultaneous testing without the inclusion of auxiliary side information. Lemma \ref{lem4} guarantees that if for each study the summary statistics are partially exchangeable and the testing procedure is symmetric then a scaled version of the evidence indices in Equation \eqref{eq:eij} are actually bonafide $e-$values.
			\begin{lemma}\label{lem4}
				Suppose for each study $j$ the following holds: (i) the summary statistics $\{X_{ij}\}_{i\in\mathcal{M}_j}$
				are partially exchangeable on $\cH_{0j}$, (ii) the testing procedure is symmetric, and (iii) controls FDR at level $\alpha_j$. Then for all $i\in\cH_{0j}$, it holds that $\EE[e_{ij}]\leq m_j/|\cH_{0j}|$ with $e_{ij}$ defined in Equation \eqref{eq:eij}. In particular, if $\pi_j\in(0,1)$ is a lower bound for  $|\cH_{0j}|/m_j$, then $\pi _je_{ij}$ is an $e-$value, i.e., $\EE[\pi _je_{ij}]\leq 1$.
				\label{prop:exchange_frequentist}
			\end{lemma}
			$e-$values are substantially more flexible than compound $e-$values. For instance, $e-$values facilitate reliable inferences for individual hypotheses while compound $e-$values are limited to simultaneous inference across a set of null hypotheses.  Furthermore, the only aggregation scheme for compound $e-$values discussed in the literature is (weighted) arithmetic mean \citep{ren2024derandomised,li2023values}. 
			In contrast, under independence bonafide $e-$values admit aggregation via multiplication, which allows evidence to ``accumulate''. Let $\mathcal N_i=\{j:\mathbb I(i\in \mathcal M_j)=1\}$ denote the set of studies that test hypothesis $H_{0i}$ with $|\mathcal N_i|=n_i$. Given a pre-determined $k\in \{1,\ldots,n_i\}$, denote $\mathcal S_{ki}$ as any $k$ element subset of $\mathcal N_i$, $i\in\mathcal M$. 
            Define $e_{i,\mathcal S_{ki}}=\prod_{j\in \mathcal{S}_{ki}}\pi_je_{ij}$. The next theorem shows that under certain conditions $e_{i,\mathcal S_{ki}}$ 
            is an $e-$value.
\begin{theorem}\label{thm3}
				Suppose (i) the conditions in Lemma \ref{lem4} hold and (ii) the summary statistics
				for the $i$-th testing problem, 
				$\{X_{ij}\}_{j\in\mathcal N_i}$ are independent
				conditional on $\theta_i$ for all $i$. Then $\mathbb{E}(e_{i,\mathcal S_{ki}})\leq 1$ for $i\in\cH_0$.
			\end{theorem}
			Theorem \ref{thm3} facilitates evidence accumulation via a product rule as we can multiply the $\{e_{ij}\}_{j\in\mathcal \mathcal S_{ki}}$'s for aggregation. However, simply multiplying them may not be ideal since if just one study in $\mathcal S_{ki}$ fails to reject $H_{0i}$ the product will be $0$. To partially over come this difficulty, we propose to use $e_i^{\tt agg*}$ as defined below for evidence aggregation.
\begin{equation}
\label{eq:eagg_alternative}
e_i^{\tt agg*} = \frac{1}{n_i}\sum_{k=1}^{n_i}\binom{n_i}{k}^{-1}\sum_{\mathcal{S}_{ki}\in \mathcal{B}_{ki}}e_{i,\mathcal S_{k,i}},
\end{equation}
where $\mathcal B_{ki}$ is the set of all $k$ element subsets of $\mathcal N_i$. The idea is to try all possible $\pi^kA_{i,\mathcal{S}_{ki}}$ and then take average. Since $\pi^kA_{i,\mathcal{S}_{ki}}$ are $e-$values, their average is also an $e-$value. The form of $e_i^{\tt agg*}$ in Equation \eqref{eq:eagg_alternative} also appears in \cite{vovk2024true} as ``U-statistics" $e-$values. We denote the procedure that applies the e-BH procedure on $(e_1^{\tt agg *},\ldots,e_m^{\tt agg *})$ as \texttt{IRT$^*$}.
In the real data analysis and numerical experiments of sections \ref{sec:realdata} and \ref{sec:sims} we take $\pi_j=0.5$ for all $j\in[d]$, as is often assumed in the literature  \citep{jin2007estimating}.
\begin{figure}[!h]
\centering
\includegraphics[width=0.65\linewidth]{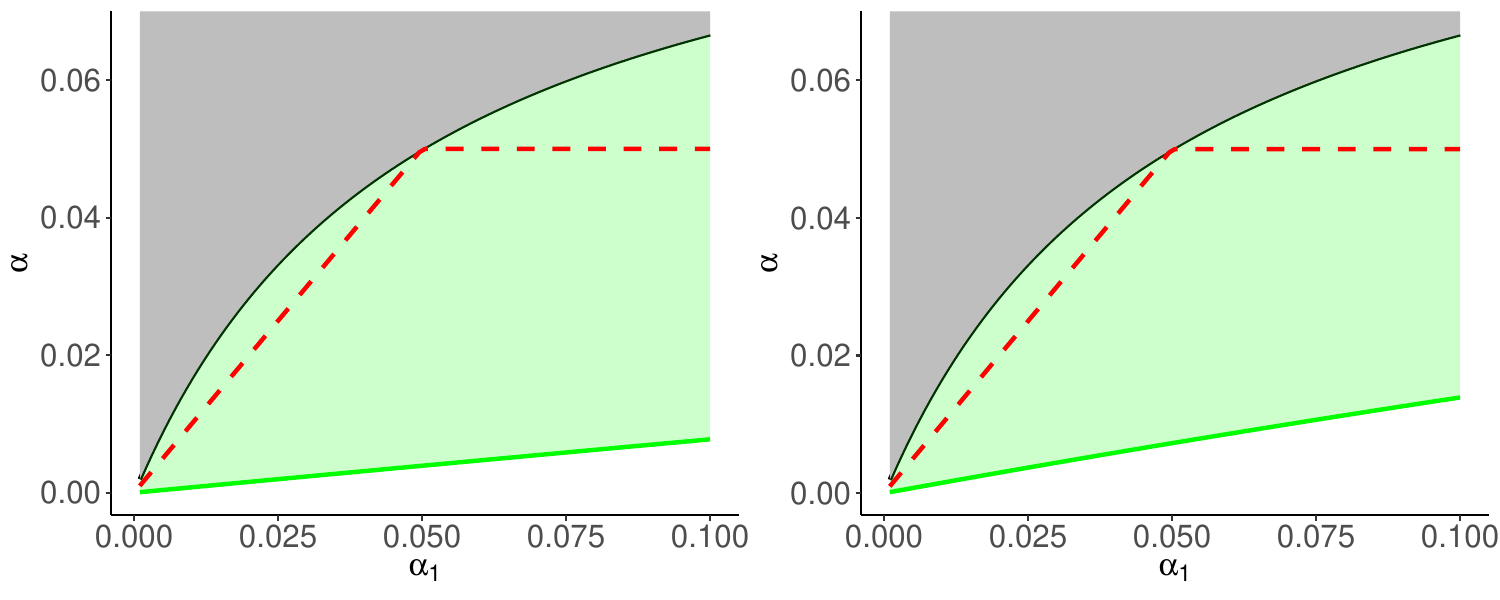}
\caption{Left: Same setting as the left panel of Figure \ref{fig:alpha_plot}. The gray shaded region depicts the overall FDR level $\alpha$ required for \texttt{IRT} to reject $H_{0i}$ while the green and the gray regions represent what value of $\alpha$ is required by \texttt{IRT$^*$}. Right: Here $d=4$ studies are testing the same set of $m=5$ null hypotheses. The first two studies are testing at level $\alpha_1$ while the remaining two at $\alpha_2$. The first two studies are using exactly the same data and both reject only the $i$-th null hypothesis at FDR level $\alpha_1$. Other two studies are independent such that Theorem \ref{thm3} holds, but both continue to reject the same $i$-th null hypothesis at FDR level $\alpha_2$. So $\delta_{ij}=1$ and $\|\bm{\delta}_j\|_0= 1$ for all $j\in[d]$. When \texttt{IRT} is used to pool the inferences, the gray shaded region depicts the overall FDR level $\alpha$ required for the e-BH procedure to reject $H_{0i}$ as $\alpha_1$ varies and $\alpha_2$ is fixed at $0.05$. The gray and the green shaded regions depict what value of $\alpha$ is required for e-BH to reject $H_{0i}$ when the hybrid scheme \texttt{IRT H} (see Remark \ref{remark: IRTH}) is used.}
\label{fig:alpha_plot_2}
\end{figure}
\begin{remark}
    \label{remark:assumptions}
 Note that from Theorem \ref{thm3}, \texttt{IRT$^*$} guarantees valid FDR control under (1) exchangeability
of the study-specific summary statistics, (2) symmetry of the study-specific decision rule and (3) independence of the summary statistics for each testing problem. If the data underlying the null hypotheses are independent, then the resulting $p-$values, which are common summary statistics, can be exchangeable, regardless of effect heterogeneity. A key aspect here is that under the null, such $p-$values should follow a Uniform(0,1) distribution. In contrast, if the data under the null hypotheses exhibit, for instance, spatial dependence then the corresponding $p-$values may not be exchangeable. In Section \ref{sec:sims_ablation} we assess the numerical performances of \texttt{IRT$^*$} when the assumptions underlying Lemma \ref{lem4} and Theorem \ref{thm3} are violated. In particular, we find that \texttt{IRT$^*$} is relatively robust to the exchangeability assumption on the
summary statistics in Lemma \ref{lem4}.
\end{remark}
\begin{remark}
\label{remark:pi}
The scaling by $\pi_j$ in Lemma \ref{lem4} is a theoretical requirement for constructing a valid e-value, but it also provides a crucial and desirable statistical calibration. The magnitude of evidence from a discovery should be calibrated by how difficult it was to make. Most modern multiple testing procedures are inherently data-adaptive, and the bar for declaring a finding significant is often lower in a dense-signal environment. In the BH procedure, for instance, a higher proportion of true signals makes the procedure more powerful, leading to a less stringent p-value threshold that satisfies the FDR criterion.
This implies that a hypothesis rejected in a dense-signal study may not have needed to achieve as small a p-value as one rejected in a sparse-signal study. It is therefore natural and statistically sound for discoveries from dense-signal settings to contribute a lower evidence weight. The scaling by $\pi_j$ in our \texttt{IRT$^*$} framework is the mechanism that performs this automatic and desirable calibration.
\end{remark}

We now return to the example considered in Section \ref{sec:alpha_choice} but instead use \texttt{IRT$^*$} to pool the inferences from the two studies. The green and the gray shaded regions in the left panel of Figure \ref{fig:alpha_plot_2} depicts the overall FDR level $\alpha$ required for the e-BH procedure to reject $H_{0i}$ as a function of $\alpha_1$ and with $\alpha_2=0.05$. Clearly, \texttt{IRT$^*$} is able to reject $H_{0i}$ even when $\alpha<\min(\alpha_1,\alpha_2)$. This represents a stark contrast to \texttt{IRT} which requires $\alpha$ to stay in the gray region to reject $H_{0i}$. The reason for this distinction is related to the fact that, courtesy Theorem \ref{thm3}, the aggregation scheme underlying \texttt{IRT$^*$} is more powerful than the arithmetic mean. Note that the calculations for the left panel directly follow from the \texttt{IRT$^*$} procedure which, in this example, rejects $H_{0i}$ at FDR level $\alpha$ if $\alpha\ge 8\alpha_1\alpha_2/(\alpha_1+\alpha_2+m)$. While this example relies on a relatively simple setting, our numerical experiments in Section \ref{sec:sims} of the supplement confirm the broader conclusion that, in general, \texttt{IRT} is not powerful when $\alpha<\min_{j\in[d]}\alpha_j$ while, under the conditions of Theorem \ref{thm3}, \texttt{IRT$^*$} is powerful in this setting. 
\begin{remark}
\label{remark: IRTH}
Motivated by the setting in the left panels of figures \ref{fig:alpha_plot} and \ref{fig:alpha_plot_2}, suppose there are $d=4$ studies, each testing the same set of $m=5$ null hypotheses. The first two studies are testing at level $\alpha_1$ while the remaining two at $\alpha_2$. Consider a setting where the first two studies are using exactly the same data and both reject only the $i$-th null hypothesis at FDR level $\alpha_1$. Other two studies are independent such that Theorem \ref{thm3} holds, but both continue to reject the same $i$-th null hypothesis at FDR level $\alpha_2$. So $\delta_{ij}=1$ and $\|\bm{\delta}_j\|_0= 1$ for all $j\in[d]$. To pool inferences across these $d$ studies, \texttt{IRT} and \texttt{IRT$^*$} can be used in conjunction. Specifically, the aggregated evidence indices from the first two studies are constructed using \texttt{IRT}, denoted $e_i^{\tt agg}$ (Equation \eqref{eq:fede}) and \texttt{IRT$^*$} is employed to aggregate the evidence indices from the remaining two studies, denoted $e_i^{\tt agg*}$ (Equation \eqref{eq:eagg_alternative}). Denote $e_i^{\tt agg, H}=(1/2)(e_i^{\tt agg}+e_i^{\tt agg*})$ as the hybrid aggregated evidence indices. Note that $\{e_i^{\tt agg, H}\}_{i=1}^{m}$ are a set of compound $e-$values under $\mathcal H_{0}$. When $\{e_i^{\tt agg, H}\}_{i=1}^{m}$ are used as inputs to the e-BH procedure, the gray and the green shaded regions in the right panel of Figure \ref{fig:alpha_plot_2} depict the overall FDR level $\alpha$ required for e-BH to reject $H_{0i}$ as a function of $\alpha_1$ and with $\alpha_2=0.05$. 
Clearly, this hybrid scheme is able to reject $H_{0i}$ even when $\alpha<\min(\alpha_1,\alpha_2)$. In contrast, the gray shaded region reveals that \texttt{IRT} has no power unless $\alpha>\min(\alpha_1,\alpha_2)$. 
				
The above example represents a practical setting where often additional information regarding data-sharing or the use of auxiliary side information for multiple testing is available for the $d$ studies. For instance, suppose prior knowledge dictates that a set of $d_1\subset[d]$ studies share data amongst themselves, while the conditions of Theorem \ref{thm3} hold for the remaining set of $d_2=[d]\setminus d_1$ studies. Denote the aggregated evidence indices across the $d_1$ studies derived from \texttt{IRT} as $\{e_i^{\tt agg, d_1}\}_{i\in\mathcal M}$ and those derived from \texttt{IRT$^*$} across the $d_2$ studies as $\{e_i^{\tt agg*, d_2}\}_{i\in\mathcal M}$. Then, the hybrid aggregated evidence indices $e_i^{\tt agg,H}=(1/d)(|d_1|e_i^{\tt agg, d_1}+|d_2|e_i^{\tt agg*, d_2})$ are a set of compound $e-$values under $\mathcal H_0$ and the e-BH procedure guarantees FDR control at level $\alpha$ when $\{e_i^{\tt agg,H}\}_{i=1}^{m}$ are used as inputs. We call this procedure \texttt{IRT H} and evaluate its empirical performance in Section \ref{sec:sims} of the supplement.
\end{remark}
\section{Proofs}
\label{app:proofs}
\subsection{Proof of Theorem \ref{thm1}}
\begin{proof}
				Based on the evidence construction in Equation \eqref{eq:eij}, we have
				\begin{align}
					\sum_{i\in \mathcal{H}_{0j}}\mathbb{E}(e_{ij})&=\dfrac{m_j}{\alpha_j}~\mathbb E\left[\dfrac{\sum_{i\in \mathcal{H}_{0j}}\delta_{ij}}{\|\bm{\delta}_j\|_0\vee 1}\right]=\dfrac{m_j}{\alpha_j}~\text{FDR}(\bm \delta_j)\leq m_j\nonumber,
				\end{align}
				where the last inequality results from the fact that study $j$ controls FDR at level $\alpha_j$.
			\end{proof}
\subsection{Proof of Theorem \ref{thm2}}
\begin{proof}
We have
\begin{align}
\sum_{i\in \mathcal{H}_{0}}\mathbb{E}[e_{i}^\texttt{agg}]&=\dfrac{1}{d}\sum_{i\in \mathcal{H}_{0}}\sum_{j=1}^{d}\Bigl\{\dfrac{m_j}{\alpha_j}~\mathbb{E}\left[\dfrac{\delta_{ij}}{\max(\|\bm\delta_j\|_0,1)}\right]~\mathbb I(i\in\mathcal M_j)+\mathbb I(i\notin\mathcal M_j)\Bigr\}\nonumber\\
&=\dfrac{1}{d}\sum_{j=1}^{d}\Bigl\{\dfrac{m_j}{\alpha_j}~{\rm FDR}(\bm \delta_j)+\sum_{i\in \mathcal{H}_{0}}\mathbb I(i\notin\mathcal M_j)\Bigr\}\nonumber\\
&\le\dfrac{1}{d}\sum_{j=1}^{d}\Bigl\{ m_j+m-m_j\Bigr\}=m,\nonumber
\end{align}
which completes the proof.
\end{proof}
			\subsection{Proof of Lemma \ref{lem4}}
			\begin{proof}
				Denote $\theta_{i,i'}=(\theta_i,\theta_{i'})$ and ${\theta}_{-i,i'}^j=\{\theta_k\}_{k\neq i,i',k\in \mathcal{M}_j}$. For all $i,i'\in\cH_{0j}$, we have
				\begin{align*}
					\mathbb{E}[e_{ij}|\theta_{i,i'}=\mathbf{0},\theta_{-i,i'}^j=\zeta]&={\int\mathbb{E}[e_{ij}|\Xb_j]\cdot\PP(\Xb_j|\theta_{i,i'}=\mathbf{0},\theta_{-i}^j=\zeta)\ \rd \Xb_j}\\
					&=\int\mathbb{E}[e_{ij}\mid\pmb{\delta}_j(\Xb_j)]\cdot{\PP(\Xb_j|\theta_{i,i'}=\mathbf{0},\theta_{-i}^j=\zeta)\ \rd \Xb_j}.
				\end{align*}
				By symmetry of the decision rule we have
				\begin{align}\label{pfeq1}
					\mathbb{E}[e_{ij}|\bm{\delta}_j(\Xb_j)]=\mathbb{E}[e_{i'j}|\bm{\delta}_j( \Psi_{i,i'}\{\Xb_{j}\})].
				\end{align}
				Furthermore, by partial exchangeability we have
				\begin{align}\label{pfeq2}
					&\PP(\Xb_j|\theta_{i,i'}=\mathbf{0},\theta_{-i,i'}^j=\zeta)=\PP(\Psi_{i,i'}\{\Xb_j\}|\theta_{i,i'}=\mathbf{0},\theta_{-i,i'}^j=\zeta).
				\end{align} 
				Thus, from equations \eqref{pfeq1} and \eqref{pfeq2} we have $\EE[e_{ij}]=\EE[e_{i'j}]$ for all $i,i'\in\cH_{0j}$. Since 
				$\sum_{i\in\cH_{0j}} \mathbb{E}[e_{ij}]\leq m_j$ as shown in Theorem \ref{thm1}, we have $\mathbb{E}[ e_{ij}]\leq m_j/|\cH_{0j}|$ for all $i\in\cH_{0j}$.
			\end{proof}		
			\subsection{Proof of Theorem \ref{thm3}}
			\begin{proof}
				Note that for $i\in\mathcal H_0$,  
$\mathbb{E}_{H_{0i}}(e_{i,\mathcal{S}_{ki}})=\prod_{j\in \mathcal{S}_{ki}}\pi_j\mathbb{E}_{H_{0i}}(e_{ij})\leq \prod_{j\in\mathcal{S}_{ki}}\pi_j\dfrac{m_j}{|\mathcal{H}_{0j}|}\le 1$.
			\end{proof}
\section{Additional technical details }
			\label{sec:discuss}
			\subsection{Connections to existing aggregation and derandomization procedures}
			\label{sec:evid_motivation}
			Leveraging $e-$values for aggregation and derandomization for specific FDR methods has been explored in literature recently \citep{ren2024derandomised,li2023values,bashari2023derandomized,zhao2024false}. In this subsection, we show that these seemly distinct $e-$values constructions can be viewed as special cases of our construction in Equation \eqref{eq:eij}
			when examined from an asymptotic perspective.
			
			A generic FDR procedure can be described abstractly as follows
			\begin{enumerate}
				\item
				\textbf{(Ranking)} Construct a suitable summary statistics $T_i$ for for each $H_{0,i}$ and rank the null hypotheses according to $T_i$.
				\item
				\textbf{(FDP Estimation)} For any given $t$ estimate the FDP of the decision rule $\pmb{\delta}(t)=\{\delta_1(t),\ldots, \delta_m(t)\}$, where $\delta_i(t)=\ind(T_i\leq t)$. Denote the estimate as $\widehat{\rm FDP}(t)$.
				\item
				\textbf{(Thresholding)} For a given target FDR level $\alpha$, define
				$
				t_\alpha=\sup\{t:\widehat{\rm FDP}(t)\leq\alpha\}
				$. Reject $H_{0,i}$ if and only if $T_i\leq t_\alpha$.

			\end{enumerate}
			Denote the above FDR procedure as $\pmb{\delta}=(\delta_{1},\ldots, \delta_m)$, where $\delta_{i}=\mathbb{I}(T_i\leq t_\alpha)$, then
			$e-$values constructed in  \cite{ren2024derandomised,li2023values,bashari2023derandomized,zhao2024false} can be written in the form of
			\begin{equation}\label{e-value}
				e_i=\dfrac{m\delta_i}{\widehat{\rm FDP}(t_\alpha)\|\pmb{\delta}\|_0}.
			\end{equation}
			Note that the denominator in Equation \eqref{e-value} can be viewed as the estimated number of false discoveries. In what follows we give two explicit examples showing that the construction in \eqref{e-value} and \eqref{eq:eij} are equivalent as $\|\pmb{\delta}\|_0\rightarrow \infty$.
			\begin{enumerate}
				\item \textbf{Asymptotic equivalence to \cite{ren2024derandomised} -} The knockoff filter \citep{barber2015controlling,barber2019knockoff} is a framework for selecting a set of covariates that are relevant for predicting a response variable $Y$ with guaranteed control of the FDR. For each covariate $X_i$, it constructs a statistic $W_i$ that is likely to be large if $X_i$ is relevant for predicting $Y$ conditional on $\{X_j\}_{j\neq i}$ and has a symmetric distribution around 0, otherwise. The knockoff filter selects $X_i$ if and only if $W_i\geq t_\alpha$ where
				\begin{equation}\label{eq:knockoff}
					t_\alpha=\inf\left\{t>0:\frac{1+\sum_{i=1}^m\II(W_i\leq-t)}{\sum_{i=1}^m\II(W_i\geq t)}\leq\alpha\right\}.
				\end{equation}
				
				\cite{ren2024derandomised} show that the following is a set of compound $e-$values.
				\begin{equation}
					e_i=\frac{m\cdot\II(W_i\geq t_\alpha)}{1+\sum_{i=1}^m\II(W_i\leq -t_\alpha)},\quad \forall i\in\cM,
					\label{eq:evid_BC}
				\end{equation}
				We now explain how Equation \eqref{eq:evid_BC} is related to Equation \eqref{eq:eij}.	Denote $\delta_i=\mathbb{I}(W_i\geq t_\alpha)$. Then $t_\alpha$ can be written as
				$
				t_\alpha=\inf\left\{t>0:1+\sum_{i=1}^m\II(W_i\leq-t)\leq\alpha\|\bm\delta\|_0\right\}.
				$
				Let $0<\breve{t}_\alpha<t_\alpha$ be such that $\sum_{i=1}^m\II(W_i\leq-\breve{t}_\alpha)=1+\sum_{i=1}^m\II(W_i\leq-t_\alpha)$. We then have
				\begin{equation*}
					\alpha\|\bm\delta\|_0=\alpha\sum_{i=1}^m\II(W_i\geq t_\alpha)\leq\alpha\sum_{i=1}^m\II(W_i\geq\breve{t}_\alpha)<1+\sum_{i=1}^m\II(W_i\leq-\breve{t}_\alpha)=2+\sum_{i=1}^m\II(W_i\leq-t_\alpha),
				\end{equation*}
				where the second inequality follows form the definition of $t_\alpha$ and $\breve{t}_\alpha$. For the decision rule $\delta_i=\mathbb{I}(W_i\geq t_\alpha)$ the evidence index defined in Equation \eqref{eq:eij} becomes ${m\II(W_i\geq t_\alpha)}/{(\alpha\|\bm\delta\|_0)}$ and
				\begin{align}
					\frac{m\II(W_i\geq t_\alpha)}{\alpha\|\bm\delta\|_0}
					&        \leq\frac{m\II(W_i\geq t_\alpha)}{1+\sum_{i=1}^m\II(W_i\leq -t_\alpha)}\leq\frac{m\II(W_i\geq t_\alpha)}{\alpha\|\bm\delta\|_0-1}\nonumber
					=\frac{m\II(W_i\geq t_\alpha)}{\alpha\|\bm\delta\|_0}\cdot\frac{\|\bm\delta\|_0}{\|\bm\delta\|_0-\alpha^{-1}},
				\end{align}
				where the first inequality again follows from the definition of $t_\alpha$ in Equation \eqref{eq:knockoff}.
				Hence, in the context of large-scale inference where $\|\bm\delta\|_0/(\|\bm\delta\|_0-\alpha^{-1})\overset{p}{\rightarrow}1$, Equation \eqref{eq:eij} and Equation \eqref{eq:evid_BC} are asymptotically equivalent. 
				\\[0.25ex]
				\item \textbf{Asymptotic equivalence to \cite{li2023values} - }
				Given null hypotheses $H_{01},\ldots H_{0m}$ and $p-$values $p_1,\ldots,p_m$, the BH procedure with target FDR level $\alpha$ rejects $H_{0i}$ if and only if 
				\begin{equation}\label{eq:bh}
					\delta_i=\II(p_i\leq t_\alpha),\text{~where~}t_\alpha=\sup\left\{t\in(0,1]:\frac{mt}{\sum_{i=1}^m\II(p_i\leq t)}\leq\alpha\right\},
				\end{equation}
				\cite{li2023values} show that the following is a set of compound $e-$values. 
				\begin{equation}
					e_i=t_\alpha^{-1}\II(p_i\leq t_\alpha),\quad \forall i\in\cM.
					\label{eq:evid_BH}
				\end{equation}
				Define
				\begin{equation}\label{eq:bh2}
					\breve{t}_\alpha=t_\alpha+\frac{1}{m}.
				\end{equation}
				We then have 
				\begin{equation}\label{eq:bh3}
					\alpha\|\pmb{\delta}\|_0=\alpha\sum_{i=1}^{m}\mathbb{I}(p_i\leq t_\alpha)\leq \alpha\sum_{i=1}^{m}\mathbb{I}(p_i\leq \breve{t}_\alpha)<m\breve{t}_\alpha=mt_\alpha+1,
				\end{equation}
				where the first and second inequality follows from the definition of $t_\alpha$ and $\breve{t}_\alpha$ in equations \eqref{eq:bh} and \eqref{eq:bh2}.
				For the BH procedure, the evidence index in Equation \eqref{eq:eij} takes the form $m\mathbb{I}(p_i\leq t_\alpha)/(\alpha \|\pmb{\delta}\|_0)$. Note that
				\begin{align*}
					\frac{m\II(p_i\leq t_\alpha)}{\alpha\|\bm\delta\|_0}
					\leq \frac{m}{m t_\alpha}\II(p_i\leq t_\alpha)\leq\frac{m\II(p_i\leq t_\alpha)}{\alpha\|\bm\delta\|_0-1}
					=\frac{m\II(p_i\leq t_\alpha)}{\alpha\|\bm\delta\|_0}\cdot\frac{\|\bm\delta\|_0}{\|\bm\delta\|_0-\alpha^{-1}},
				\end{align*}
				where the first inequality follows from Equation \eqref{eq:bh} and the second inequality follows from Equation \eqref{eq:bh3}. For large-scale inference problems, where $\|\bm\delta\|_0/(\|\bm\delta\|_0-\alpha^{-1})\overset{p}{\rightarrow}1$, Equation \eqref{eq:eij} and Equation \eqref{eq:evid_BH} are asymptotically equivalent. 
			\end{enumerate}
			\begin{remark}
				The ranking statistic employed in \cite{dai2023false,dai2023scale} for derandomization can be expressed as
				$ \delta_i/(\widehat{\rm FDP}(t_\alpha)\|\pmb{\delta}\|_0).
				$
				This formulation yields the same ranking as \eqref{e-value}. However, \cite{dai2023false,dai2023scale} do not use $e$-BH for aggregation.
			\end{remark}
			\subsection{Asymptotic FDR control} 
			\label{sec:agent_fdr}
			A key requirement for the validity of the \texttt{IRT} procedure is that the study-specific multiple testing procedure controls FDR at their pre-specified level $\alpha_j$. Theorems \ref{thm2} and \ref{thm3} implicitly assume that such an FDR control holds for finite samples, i.e. $\mathbb E[\sum_{i\in \mathcal{H}_{0j}}\delta_{ij}/\|\bm \delta_j\|_0\vee 1]\le \alpha_j$ for all $j\in[d]$. In reality, however, for some studies their FDR control may be asymptotic in $m_j$. In such a scenario, the \texttt{IRT} procedure guarantees FDR control at level $\alpha$ as $m_j\to\infty$. We summarize the above discussion in the following proposition.
			\begin{proposition}\label{prop2}
				Suppose study $j$ controls FDR at level $\alpha_j$ asymptotically, i.e., ${\rm FDR}(\bm \delta_j)\leq\alpha_j+o_p(1)$. Then, \texttt{IRT} controls FDR at level $\alpha$ asymptotically.
			\end{proposition}
			\begin{proof}
				We first establish that $\eb^{\tt agg}$ in Equation \eqref{eq:fede} is a set of compound $e-$values asymptotically. Let $e_{i}^{\tt agg}$ be as defined in Equation \eqref{eq:fede}.
				Akin to the proof of Theorem \ref{thm2}, we have
\begin{align}
\sum_{i\in \mathcal{H}_{0}}\mathbb{E}[e_{i}^\texttt{agg}]&=\dfrac{1}{d}\sum_{i\in \mathcal{H}_{0}}\sum_{j=1}^{d}\Bigl\{\dfrac{m_j}{\alpha_j}~\mathbb{E}\left[\dfrac{\delta_{ij}}{\max(\|\bm\delta_j\|_0,1)}\right]~\mathbb I(i\in\mathcal M_j)+\mathbb I(i\notin\mathcal M_j)\Bigr\}\nonumber\\
&=\dfrac{1}{d}\sum_{j=1}^{d}\Bigl\{\dfrac{m_j}{\alpha_j}~{\rm FDR}(\bm \delta_j)+\sum_{i\in \mathcal{H}_{0}}\mathbb I(i\notin\mathcal M_j)\Bigr\}\nonumber\\
&\le\dfrac{1}{d}\sum_{j=1}^{d}\Bigl\{ m_j(1+o_p(1))+m-m_j\Bigr\}\leq m(1+o_p(1)).\nonumber
\end{align}

				Next, we consider $e_i^{\tt agg *}$ from Equation \eqref{eq:eagg_alternative}. Similarly, following the same arguments in Lemma \ref{lem4}, we can show that $\mathbb{E}[e_{ij}]\leq (1/\pi_j)(1+o_p(1))$. Thus, using the same notation as in Theorem \ref{thm3}, we have $
\mathbb{E}[\pi^kA_{i,\mathcal{S}_{ki}}]\leq 1+o_p(1)$. It follows that $\eb^{\tt agg*}=\{e_i^{\tt agg *}\}_{i\in\mathcal M}$ is also a set of asymptotic compound $e-$values. Let $\bm{\delta}=\{\delta_1,\ldots,\delta_m\}$ be the decision of e-BH applied on a set of asymptotic compound $e-$values $\eb^{\tt agg}$. Note that $\delta_i=1$ indicates that $ e_{i}^\texttt{agg} \geq m/\{\alpha \max(\|\bm{\delta}\|_0, 1)\}$ based on the decision rule of e-BH procedure. Thus, it holds that
\begin{align*}
\text{FDR}(\bm{\delta})&=\EE\left[\sum_{i=1}^{m} \dfrac{\mathbb{I}(\delta_i=1,\theta_i=0)}{\|\bm{\delta}\|_0\vee 1}\right]\leq\EE\left[\sum_{i=1}^{m}\dfrac{\alpha }{m}\cdot e_{i}^\texttt{agg}\mathbb{I}(\delta_i=1,\theta_i=0)\right]\\
					&\leq\frac{\alpha}{m}\cdot\EE\left[\sum_{i=1}^{m}e_{i}^{\tt agg}\mathbb{I}(\theta_i=0)\right]=\frac{\alpha}{m} \cdot m(1+o_p(1))=\alpha+o_p(1).\end{align*}
\end{proof}	
\section{{IRT} for alternative forms of Type I error control}
\label{sec:alternative_typeI}
\subsection{\texttt{IRT} for $k$-Family-wise error rate ($k$-FWER) control}
			We consider a setting where study $j$ controls the $k_j-$FWER at level $\alpha_j$, i.e.
			\begin{equation}
				\label{eq:k-fwer}  
				\PP\Big(\sum_{i\in\mathcal{H}_{0j}}\delta_{ij}\geq k_j\Big)\leq\alpha_j,\quad\forall j\in[d].
			\end{equation}
			Under this setting, the three steps of the \texttt{IRT} procedure are as follows.
			\\[1ex]
			\noindent\textbf{Evidence construction.} Suppose $k_j>1$. We consider the following evidence index,
			\begin{equation}
				\label{eq1}
				e_{ij}=\dfrac{m_j\delta_{ij}}{
					\max\{ c_1\alpha_j\|\bm{\delta}_j\|_0, c_2(k_j-1), \alpha_j\}},\ \text{where}~c_1,c_2> 0 ~\text{and}~ \frac{1}{c_1}+\frac{1}{c_2}=1,\quad\forall i\in\mathcal{M}_j.
			\end{equation}
			If study $j$ satisfies Equation \eqref{eq:k-fwer} then the evidence indices in Equation \eqref{eq1} are compound $e-$values. To see this, let $V_j$ denote the number of false rejections made by study $j$. Then,
			\begin{align*}
				&\mathbb{E}\left[ \sum_{i\in \mathcal H_{0j}}\dfrac{m_j\delta_{ij}}{
					\max\{ c_1\alpha_j\|\bm{\delta}_j\|_0, c_2(k_j-1), \alpha_j\}}  \right]\nonumber\\
				&=\mathbb{E}\left[ \sum_{i\in \mathcal H_{0j}}\dfrac{m_j\delta_{ij}}{
					\max\{ c_1\alpha_j\|\bm{\delta}_j\|_0, c_2(k_j-1), \alpha_j\}}  \bigg| V_j>k_j\right]\mathbb{P}(V_j\geq k_j)\nonumber\\
				&+ \mathbb{E}\left[ \sum_{i\in \mathcal H_{0j}}\dfrac{m_j\delta_{ij}}{
					\max\{ c_1\alpha_j\|\bm{\delta}_j\|_0, c_2(k_j-1), \alpha_j\}}  \bigg| V_j<k_j\right]\mathbb{P}(V_j< k_j)\nonumber\\
				&\leq \dfrac{m_j}{c_1\alpha_j}\alpha_j+ \frac{m_j}{c_2}\cdot 1=m_j\left(\frac{1}{c_1}+\frac{1}{c_2}\right)=m_j\label{eq5}
			\end{align*}
			We recommend choosing $c_1=(2k_j-1)/k_j, c_2=(2k_j-1)/(k_j-1)$. The rationale is that when $\alpha_j\|\bm{\delta}_j\|_0=k_j$ (as both are estimates of the number of false positives), $c_1=(2k_j-1)/k_j, c_2=(2k_j-1)/(k_j-1)$ is the  solution to the following optimization problem 
			\begin{equation*}
				\text{minimize}\ \ \ \\ \max\{ c_1\alpha_j\|\bm{\delta}_j\|_0, c_2(k_j-1)\}\quad\text{subject to}\quad c_1,c_2> 0 ~\text{and}~ \frac{1}{c_1}+\frac{1}{c_2}=1.
			\end{equation*}
			An important special case is $k_j=1$. For this setting, we fix $c_1=1$ in Equation \eqref{eq1} and recover the evidence index proposed in Equation \eqref{eq:eij} for integrative FDR control. This is not surprising since any method that controls FWER at level $\alpha_j$ also controls FDR at level $\alpha_j$. 
			
			We note that if additional information, such as what procedure study $j$ used to control $k_j$-FWER, is available then it becomes feasible to devise more powerful compound $e-$values. For example, if Bonferroni procedure is used (i.e. $H_{0i}$ is rejected by study $j$ if and only if its $p-$value is $\leq k_j\alpha_j/m_j$), then we can verify that 
			$
			e_{ij}={m_j\delta_{ij}}/{\alpha_jk_j}$ for all $i\in\mathcal{M}_j$, is also a compound $e-$value. To see this, observe that 
			$
			\mathbb{E}[\sum_{i\in \mathcal H_{0j}}e_{ij}]\leq {m_j}(k_j\alpha_j)^{-1}	\mathbb{E}[\sum_{i\in \mathcal H_{0j}}\delta_{ij}]= {m_j}{(k_j\alpha_j)^{-1}}\sum_{i\in \mathcal H_{0j}}\mathbb{P}(\delta_{ij}=1)\leq {m_j}({k_j\alpha_j})^{-1} \sum_{i\in \mathcal H_{0j}}{k_j\alpha_j}/{m_j}\leq m_j.
			$\\[1.25ex]
			\noindent\textbf{Evidence aggregation.} Since the evidence indices in Equation \eqref{eq1} are compound $e-$values, Equation \eqref{eq:fede} provides the evidence aggregation scheme in this setting and Theorem \ref{thm2} guarantees that these aggregated evidences continue to be compound $e-$values associated with $\mathcal H_0$. Furthermore, if the conditions of Theorem \ref{thm3} hold then Equation \eqref{eq:eagg_alternative} represents the aggregated evidence indices.
			\\[1.25ex]
			\noindent\textbf{$k$-FWER control.} Given the compound $e-$values $\eb^{\tt agg}=\{e_1^{\tt agg},\ldots,e_m^{\tt agg}\}$ from Step 2 above, \texttt{IRT} rejects $H_{0i}$ if and only if $e_i^{\tt agg}\ge m/(\alpha k)$. This procedure controls the $k$-FWER at level $\alpha$ since 
			\begin{align*}
				\mathbb{P}\left(\sum_{i\in \mathcal{H}_0} \mathbb{I}\Big(e_i^{\tt agg}\geq \frac{m}{\alpha k}\Big)\geq k\right)\leq  \dfrac{1}{k}\mathbb{E}\left[  \sum_{i\in \mathcal{H}_0}\mathbb{I}\Big(e_i^{\tt agg}\geq \dfrac{m}{\alpha k}\Big)\right]\leq \dfrac{1}{k}\mathbb{E} \left[\sum_{i\in \mathcal{H}_0}\dfrac{e_i^{\tt agg}\alpha k}{m}\right]\leq \alpha.
			\end{align*}
			\subsection{\texttt{IRT} for Per-family error rate (PFER) control}
			Suppose study $j$'s testing procedure controls PFER at level $k_j$, that is $\mathbb{E}\left[\sum_{i\in \mathcal H_{0j}}\delta_{ij}\right]\leq k_j$ for 
			$j\in[d]$. 
			\\[1ex]
			\noindent\textbf{Evidence construction.} We consider the evidence index
			\begin{equation}
				\label{eq:eij_pfer}
				e_{ij}=\dfrac{m_j\delta_{ij}}{k_j},\quad \forall i\in \mathcal{M}_j.  
			\end{equation}
			It is then straightforward to check that $\eb_j=\{e_{ij}\}_{i\in\mathcal{M}_j}$ are a set of compound $e-$value associated with  $\cH_{0j}$.
			\\[1.25ex]
			\noindent\textbf{Evidence aggregation.} The evidence indices in Equation \eqref{eq:eij_pfer} are compound $e-$values. Therefore, Equation \eqref{eq:fede} continues to provide the evidence aggregation scheme in this setting and Theorem \ref{thm2} guarantees that these aggregated evidences are compound $e-$values associated with $\mathcal H_0$. Furthermore, if the conditions of Theorem \ref{thm3} hold then Equation \eqref{eq:eagg_alternative} represents the aggregated evidence indices.
			\\[1.25ex]
			\noindent\textbf{$k$-PFER control.} Given the Genralized $e-$values $\eb^{\tt agg}=\{e_1^{\tt agg},\ldots,e_m^{\tt agg}\}$ from Step 2 above, \texttt{IRT} rejects $H_{0i}$ if and only if $e_i^{\tt agg}\ge m/k$. This procedure controls the PFER at level $k$ since 
				$			\mathbb{E}[\sum_{i\in \mathcal{H}_0}\mathbb{I}\Big(e_i\geq m/k\Big)]\leq 	\mathbb{E}[\sum_{i\in \mathcal{H}_0}\mathbb{I}\Big(e_ik/m\geq 1\Big)]\leq 	\mathbb{E}[\sum_{i\in \mathcal{H}_0}(e_ik/m)]\leq {k}{m^{-1}}\mathbb{E}[\sum_{i\in \mathcal{H}_0}e_i]\leq k$.
			\begin{remark}
				\label{rem:irt_type1}
				The \texttt{IRT} framework can be used for integrative inference even when studies employ different type I error control metrics. Suppose, for instance, that $d_1$ studies control $k-$FWER, $d_2$ control PFER and the remaining $d_3$ studies control FDR at desired levels, where $d_i\subset[d], i=1,2,3$, $\cap_{i=1}^{3}d_i=\emptyset$ and $\cup_{i=1}^{3}d_i=[d]$. Denote $e_{ij}^{d_1},j\in d_1,$ as the evidence indices for the $d_1$ studies from Equation \eqref{eq1}. Similarly, $e_{ij}^{d_2}$ and $e_{ij}^{d_3}$ denote, respectively, the evidence indices for studies $d_2$ and $d_3$ from equations \eqref{eq:eij_pfer} and \eqref{eq:eij}. Then $e_i=m\{\sum_{r=1}^3(\sum_{j\in d_r}e_{ij}^{d_r}/\sum_{j\in d_r}m_j)\},~i\in\mathcal M$, are a set of compound $e-$values under $\mathcal H_0$, and the e-BH procedure can be applied to $\{e_i\}_{i\in\mathcal M}$ if, for example, FDR control is the goal. Furthermore, both \texttt{IRT$^*$} and \texttt{IRT H} are also applicable in this setting if prior knowledge regarding data-sharing or the use of auxiliary side information for multiple testing is available for the $d$ studies.
			\end{remark}
\section{Additional numerical experiments}
\label{sec:sims}
\subsection{$d_1$ studies report $p-$values while $d_2$ studies report $\{\mathcal D_j\}_{j\in[d_2]}$}
We continue the discussion from Section \ref{sec:examples} and illustrate the performance of \texttt{IRT} on two additional settings.
\begin{example}
\label{ex:3}
\begin{figure}[!t]
\centering
\includegraphics[width=0.8\linewidth]{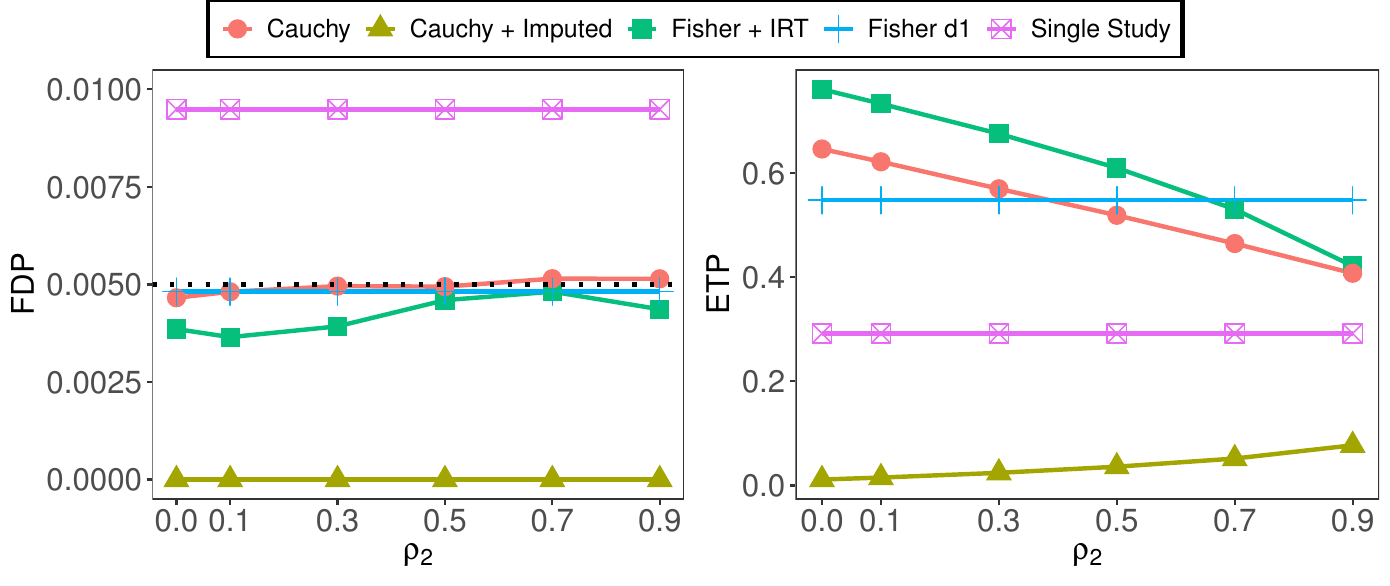}
\caption{FDP and ETP comparison for Example \ref{ex:3}.}
\label{fig:meta_blocdep_3}
\end{figure}
This is another setting with asymmetric dependence, where the $p-$value studies are independent $(\rho_1=0)$ while the binary-decision studies are dependent (with correlation $\rho_2$). We continue to borrow the setting of Example \ref{ex:1} but fix $d=15, d_1=2$ and and vary $\rho_2\in\{0,0.1,0.3,0.5,0.7,0.9\}$. We also introduce two new procedures: (i) \textbf{Fisher $d_1$}, which pools $p-$values from the first $d_1$ studies using Fisher's method \citep{fisher1948}, and then applies the BH correction, and (ii) \textbf{Fisher + IRT}, which is similar in spirit to \texttt{Cauchy + IRT} but pools the $p-$values from the first $d_1$ studies using Fisher's method. Figure \ref{fig:meta_blocdep_3} reports the average
FDP and ETP for various methods across 2000 Monte-Carlo repetitions \footnote{Whenever $\alpha<0.01$ in our numerical experiments, we set the number of Monte-Carlo repetitions to 2000 to improve the precision of the simulation estimates. Otherwise, we set it to 500.}. We continue to find that \texttt{Fisher + IRT} dominates all other methods in terms of power. In contrast, inferences from \texttt{Cauchy + Imputed} and from a single study are among the least powerful.
\end{example}
\begin{example}
\label{ex:4}
This is a setting where all $d$ studies are independent $(\rho_1=\rho_2=0)$. We fix $d=15, d_1=2$ and sample $\mu_i$ from $\pi_{0}~\delta_{\{0\}}+0.5(1-\pi_0)~\cN(3,1)+0.5(1-\pi_0)~\cN(-3,1)$. Table \ref{tab:example_4} reports the FDP and ETP comparisons for the various competing methods when $\pi_0\in\{0.5,0.8,0.95\}$. Here \textbf{Fisher + Imputed} is similar to \texttt{Cauchy + Imputed} but pools the $p-$values from the $d$ studies using Fisher's method. We find that \texttt{Fisher} dominates all methods in power, which is expected. Importantly, \texttt{Fisher + IRT} is the next best, illustrating the benefit of \texttt{IRT} in synthesizing inferences using binary decisions.
\begin{table}[!t]
\caption{FDP and ETP comparison for Example \ref{ex:4}.}
\label{tab:example_4}
\centering
\begin{tabular}{lcccccc}
\toprule
& \multicolumn{2}{c}{$0.5$}                              & \multicolumn{2}{c}{$0.8$} & \multicolumn{2}{c}{$0.95$}   \\
\toprule
Method ($\alpha=0.5\%$)           & FDP                  & ETP                           & FDP                  & ETP   & \multicolumn{1}{c}{FDP} & \multicolumn{1}{c}{ETP} \\
\hline
Fisher  & 0.0025 & 0.961 & 0.004 & 0.955 &     0.0046   &  0.946                       \\
Fisher + IRT     & 0.0025 & 0.885 & 0.0037 & 0.845 &   0.004                      &   0.763                 \\
Fisher + Imputed & $0.000$ & 0.724 & $0.000$ & 0.650 &                  0.000      &   0.538       \\
Fisher $d_1$        & 0.0025 & 0.691 & 0.0038 & 0.632 &                  0.0047       &  0.550   \\
Single Study ($\alpha_1=1\%)$     & 0.005 & 0.493  & 0.0079 &  0.403  &  0.0092    & 0.290\\
\hline
\end{tabular}
\end{table}
\end{example}
\subsection{All $d$ studies report decision sequences $\{\mathcal D_j\}_{j\in[d]}$.}
\label{sec:more_sims}
We assess the empirical performances of \texttt{IRT, IRT$^*$} and \texttt{IRT H} on simulated data when all studies report binary decisions. We consider seven simulation scenarios with $m=1000$ and test $H_{0i}:\mu_i=0~vs~H_{1i}:\mu_i\ne 0$, where 
$\mu_i\stackrel{\rm i.i.d.}{\sim}0.8\cdot\delta_{\{0\}}+0.1\cdot\cN(3,1)+0.1\cdot\cN(-3,1)$, and $\delta_{\{a\}}$ denotes a point mass at $a$. In each scenario, study $j$ uses data $X_{ij}$, to be specified subsequently, to conduct $m_j$ tests and reports the corresponding decisions $\bm \delta_j$ obtained from the BH procedure with control level $\alpha_j$. For \texttt{IRT H}, we use the following scheme across all our simulation settings: the \texttt{IRT} aggregation scheme (Equation \eqref{eq:fede}) is employed for the first $d_1=\lfloor d/2\rfloor$ studies and the \texttt{IRT$^*$} aggregation scheme (Equation \eqref{eq:eagg_alternative}) is used for the remaining $d_2=d-d_1$ studies. 
\begin{figure}[!h]
\centering
\includegraphics[width=0.8\textwidth]{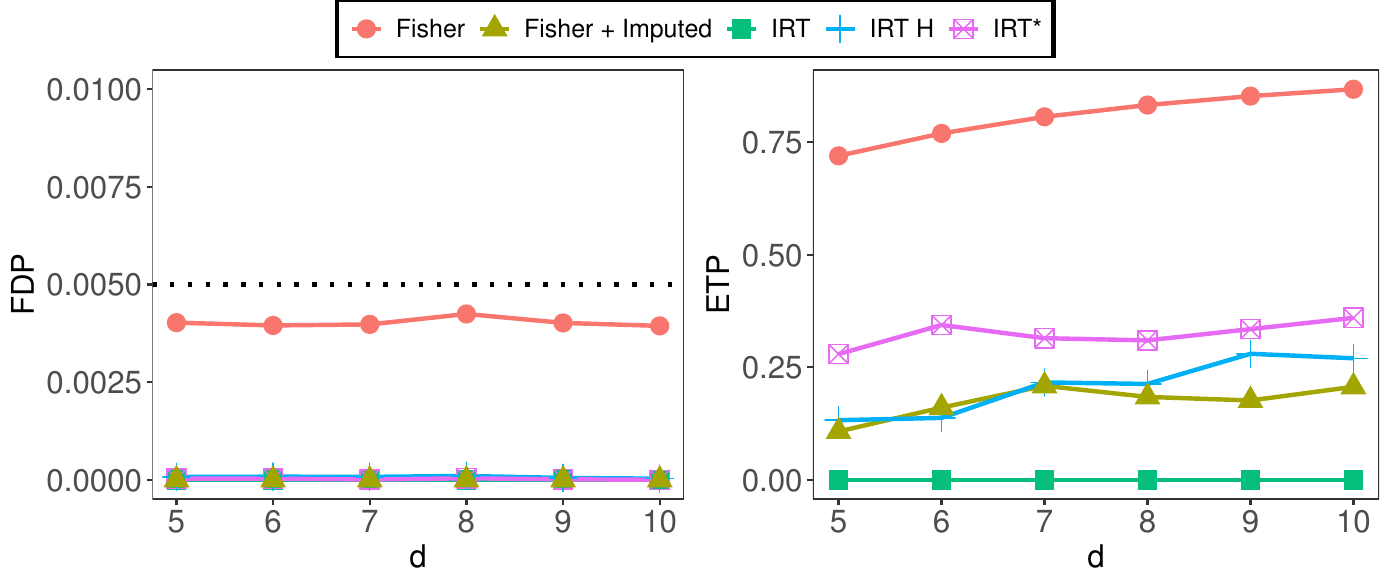}
\caption{FDP and ETP comparisons for Scenario 1.}
\label{fig:scenario_1}
\end{figure}
\\[1ex]
\noindent \textbf{Scenario 1 (Independent studies).} We begin by considering $d$ independent studies. Specifically, we let $X_{ij}\mid\mu_i,\sigma_j\stackrel{\rm ind.}{\sim}\cN(\mu_i,\sigma_j^2),~\sigma_j\stackrel{\rm i.i.d.}{\sim}{\rm Unif}(0.75,2),m_j=m$, $\alpha_j=0.01$, $\alpha=0.005$ and vary $d$ from $5$ to $10$. The empirical performances of \texttt{IRT} and its derivatives are compared against two alternative procedures: (\romannumeral1) \texttt{Fisher}, which pools the study specific $p-$values using Fisher's method \citep{fisher1948}, and then applies the BH procedure on the pooled $p-$value sequence for FDR control, and (\romannumeral2) the \texttt{Fisher + Imputed} procedure which first imputes the $p-$values using Equation \eqref{eq:conserv_pval}, pools the study specific $p-$values using Fisher's method, and then applies the BH procedure on the pooled $p-$value sequence. {When the null distribution of the test statistic is correctly specified and the corresponding $p-$values are independent, we expect \texttt{Fisher} to exhibit higher power than \texttt{IRT} and its derivatives. Nevertheless, in such settings \texttt{Fisher} provides a practical benchmark for assessing the empirical performances of \texttt{IRT, IRT$^*$} and \texttt{IRT H}, which rely only the binary decision sequences $\bm \delta_j$.}

Figure \ref{fig:scenario_1} presents the average FDP and the ETP of various methods. We make several observations. First, while all methods control the FDR at $\alpha$, \texttt{Fisher}, unsurprisingly, has the highest power across all values of $d$ and is followed by \texttt{IRT$^*$}. Second, \texttt{IRT H} is more powerful than \texttt{IRT}. In fact, the latter exhibits no power since $\alpha<\alpha_j$ for all $j\in[d]$ in this setting, further reinforcing the discussion in Section \ref{sec:alpha_choice} and Remark \ref{remark: IRTH}. However, this is not the case when the studies are correlated and $\alpha>\max_{j\in[d]}\alpha_j$, as scenarios 3 and 4 demonstrate. Third, \texttt{IRT$^*$} is more powerful than \texttt{Fisher + Imputed}, which employs valid, but conservative, $p-$values. Finally, \texttt{IRT$^*$} is more powerful than \texttt{IRT H}. This is expected since in this setting the conditions of Theorem \ref{thm3} hold and the aggregation scheme of Equation \eqref{eq:eagg_alternative} results in a more powerful procedure than \texttt{IRT H}.
\begin{figure}[!h]
\centering
\includegraphics[width=0.8\textwidth]{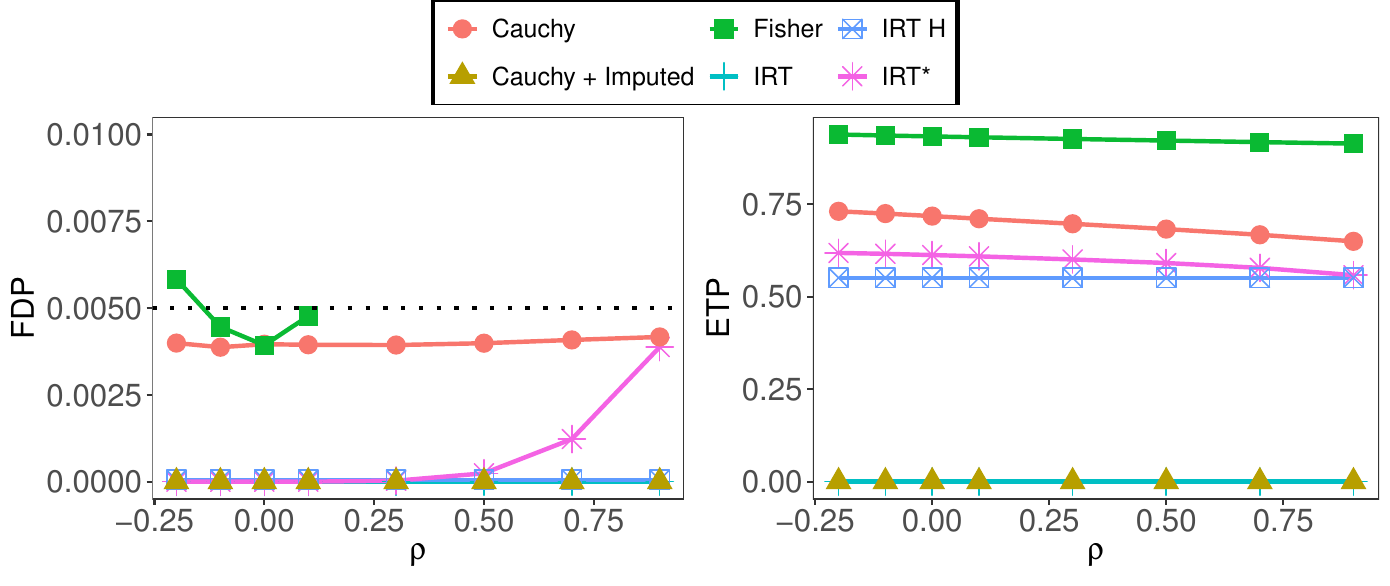}
\caption{FDP and ETP comparison for Scenario 2.}
\label{fig:scenario_2}
\end{figure}
\\[1ex]
\noindent \textbf{Scenario 2 (Correlated studies I).} The data are generated according to Scenario 1 with $\sigma_j=1,d=10,\alpha_j=0.01,\alpha=0.005$ but we introduce correlation across the first $d_1=\lfloor d/2\rfloor$ studies. In particular, we generate $(X_{i1},\ldots, X_{i\lfloor d/2\rfloor})$ from a multivariate normal distribution and set $\text{Corr}(X_{ij},X_{ik})=\rho$ for all $(j,k)\in\{1,\ldots,\lfloor d/2\rfloor\}$, $j\ne k$, where $\rho\in \{-0.1,0,0.1,0.3,0.5,0.7,0.9\}$. Along with \texttt{Fisher}, we include \texttt{Cauchy} and \texttt{Cauchy + Imputed} in our comparisons. The former pools the study specific $p-$values using the Cauchy combination test statistic \citep{liu2020cauchy}, and then applies the BH procedure on the pooled $p-$value sequence for FDR control while the latter is similar to \texttt{Fisher + Imputed} but pools the imputed $p-$values using the Cauchy combination test statistic.

Figure \ref{fig:scenario_2} reports the average FDP and the ETP for various methods. In this scenario, the first $d_1$ test statistics and the corresponding $p-$values for each hypothesis are not independent unless $\rho=0$, thus violating the conditions of Theorem \ref{thm3}. Consequently, \texttt{IRT$^*$} and \texttt{Fisher} no longer guarantee FDR control. Indeed, the left panel of Figure \ref{fig:scenario_2} reveals that \texttt{Fisher} fails to control the FDR at $0.5\%$ for large $\rho$ and therefore does not appear in the plot for some values of $\rho$. While \texttt{IRT$^*$} appears to control the FDR, it has no theoretical support for FDR guarantee in this setting. From the right panel, we find that \texttt{IRT H} is the most powerful procedure in this setting that also provably controls FDR.
\begin{figure}[!h]
\centering
\includegraphics[width=0.8\textwidth]{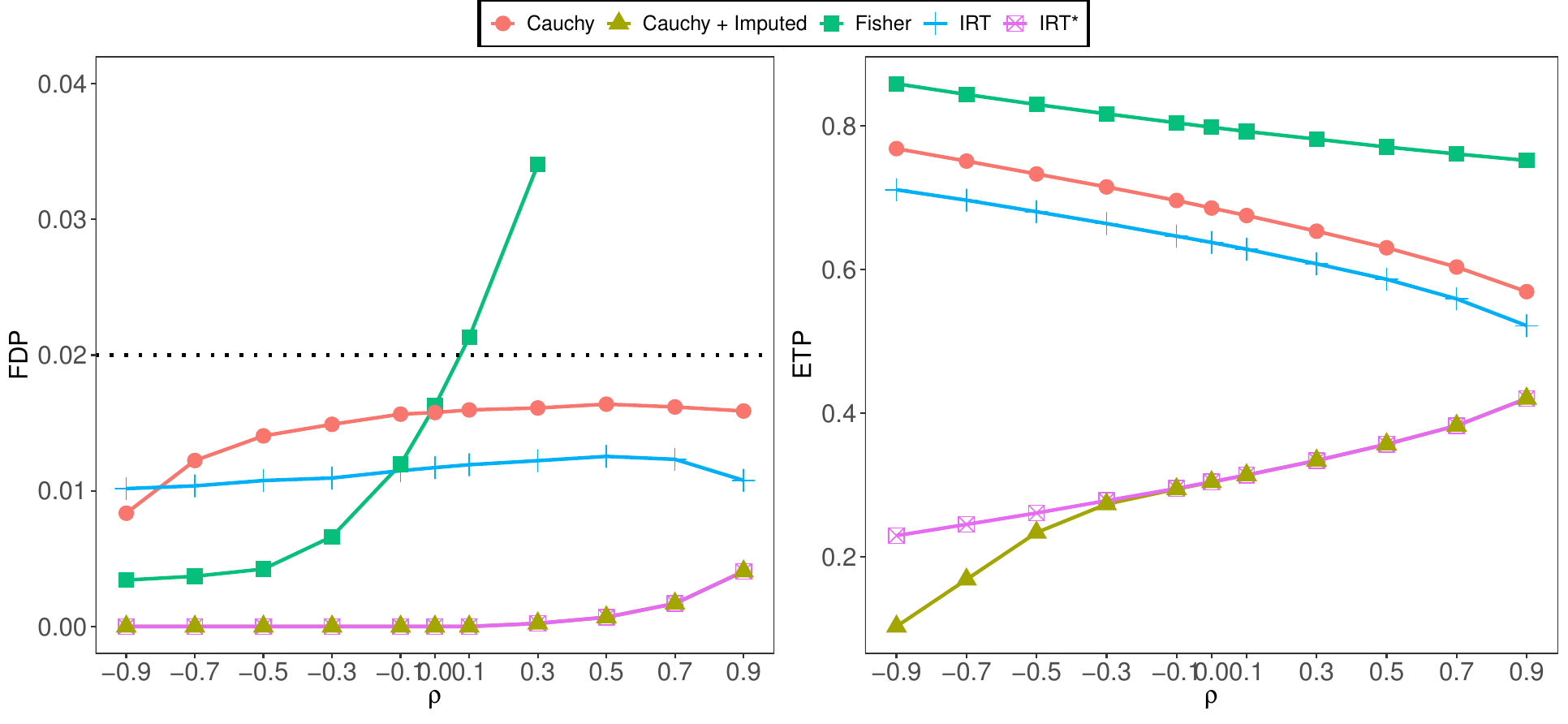}
\caption{FDP and ETP comparison for Scenario 3.}
\label{fig:scenario_3}
\end{figure}
\\[1ex]
\noindent \textbf{Scenario 3 (Correlated studies II).} In this scenario we allow all $d$ studies to be dependent and evaluate various methods using one-sided $p-$values because two-sided $p-$values from Gaussian distributions only allow positive dependence among $p-$values across studies, regardless of the correlation parameter $\rho$. Specifically, we test $H_{0i}:\mu_i=0~\text{vs}~H_{1i}:\mu_i>0$ where $\mu_i\stackrel{\rm i.i.d.}{\sim}0.8\cdot\delta_{\{0\}}+0.2\cdot\cN(3,1)$. We continue to borrow other settings from Scenario 1 with $\sigma_j=1,~d=2,~\alpha_j=0.01,~\alpha=0.02$ and let $\text{Corr}(X_{ij},X_{ik})=\rho$ for all $j\ne k$. Figure \ref{fig:scenario_3} reports the average FDP and the ETP for various methods. In this scenario, the $d$ test statistics and the corresponding $p-$values for each hypothesis are not independent unless $\rho=0$. Thus, \texttt{IRT$^*$} has no theoretical support for FDR control at level $\alpha$. Moreover, while \texttt{Fisher} does not control the FDR at $2\%$ for $\rho>0$, we find that it controls the FDR whenever $\rho<=0$, thus demonstrating less sensitivity to negative dependence. Finally, \texttt{Cauchy} and \texttt{IRT} are the next best powerful procedures in this setting that also provably control FDR. 
\\[1ex]
\noindent \textbf{Scenario 4 (Correlated studies and dependent test statistics).} We return to two-sided $p-$values in this scenario. We generate the data from Scenario 1 with $\sigma_j=1,\alpha_j=0.01,\alpha=0.02$ and introduce correlation across the studies as well as the test statistics. In particular, we let $\text{Corr}(X_{ij},X_{ik})=0.7,~j\ne k$, $\text{Corr}(X_{ij},X_{rj})=0.5,i\ne r$ and rely on the following scheme to simulate this data. For $i\in [m]$ and $j\in[d]$, sample $Y_{ij}\stackrel{i.i.d}{\sim} N(0,1)$ and denote $\bm Y$ as the $m\times d$ matrix with entries $Y_{ij}$. Let $\bm A=(1-0.5)\bm I_m+0.5\bm 1_m\bm 1_m^T$, $\bm B=(1-0.7)\bm I_d+0.7\bm 1_d\bm 1_d^T$ and suppose $\bm A=\bm U\bm U^T$, $\bm B=\bm V\bm V^T$ denote the Cholesky decompositions of $\bm A$ and $\bm B$. Then $\bm X=\bm \mu\bigotimes\bm 1_d^T+\bm U^T\bm Y\bm V$ has matrix Normal distribution, denoted $MN(\bm \mu\bigotimes\bm 1_d^T,\bm A,\bm B)$, where $\bigotimes$ denotes the usual Kronecker product, $\bm \mu\bigotimes\bm 1_d^T$ is the location and $\bm A,~\bm B$ are the scales. In particular, this implies $\text{Corr}(X_{ij},X_{rj})=0.5,~r\ne i$ and $\text{Corr}(X_{ij},X_{ik})=0.7,~j\ne k$.

\begin{figure}[!h]
\centering
\includegraphics[width=0.8\textwidth]{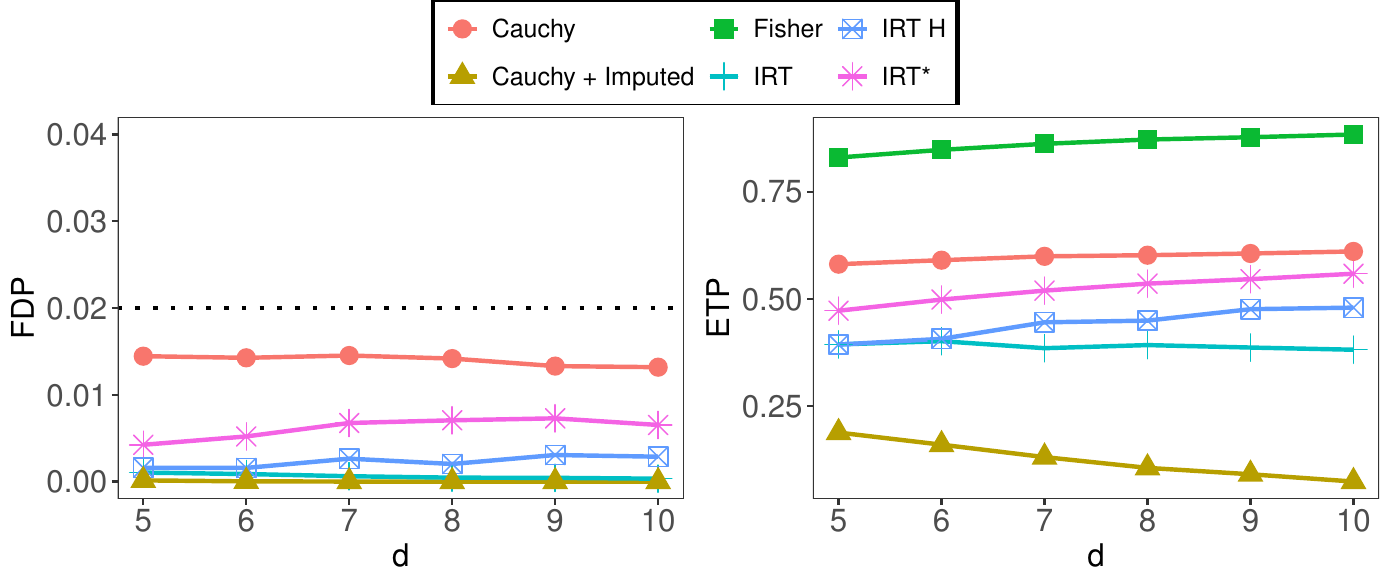}
\caption{FDP and ETP comparison for Scenario 4.}
\label{fig:scenario_4}
\end{figure}

Figure \ref{fig:scenario_4} reports the average FDP and the ETP for various methods as $d$ varies from $5$ to $10$. In this setting too \texttt{IRT$^*$}, \texttt{IRT H} and \texttt{Fisher} no longer enjoy theoretical guarantees for FDR control. We see a similar pattern as in Figure \ref{fig:scenario_2} where \texttt{Fisher} fails to control the FDR at $2\%$ whenever $\rho>0$ while \texttt{IRT} provably controls the FDR and is substantially more powerful than \texttt{Cauchy + Imputed}. 
\begin{figure}[!h]
\centering
\includegraphics[width=0.8\textwidth]{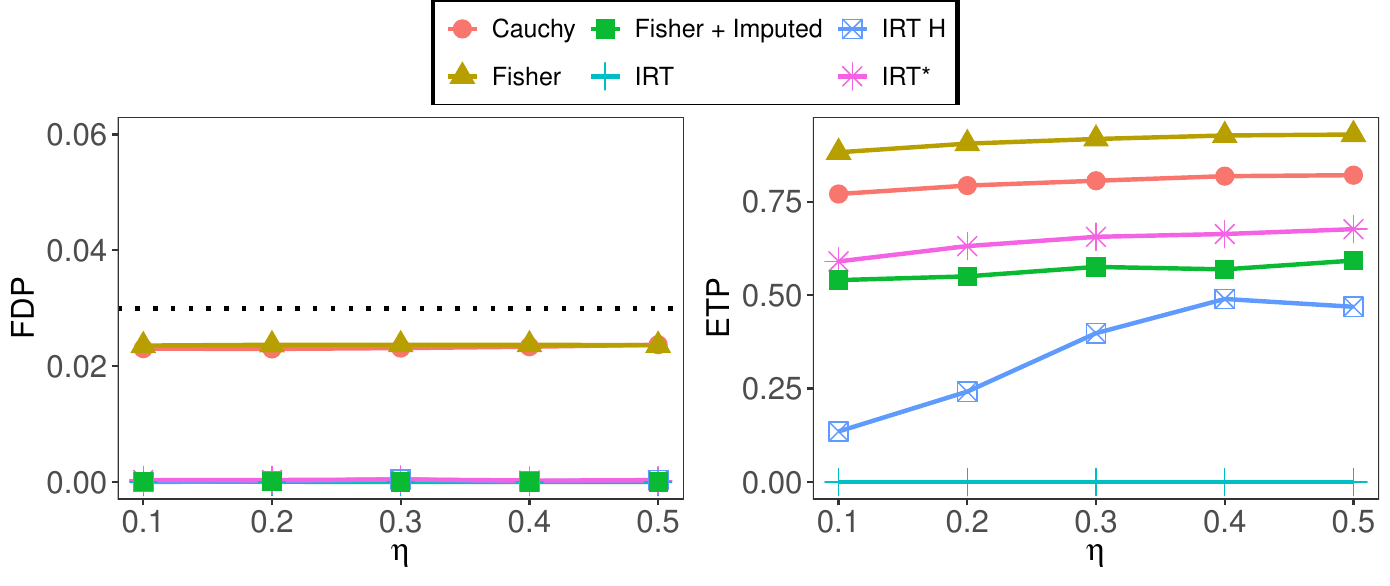}
\caption{FDP and ETP comparison for Scenario 5.}
\label{fig:scenario_5}
\end{figure}
\\[1ex]
\noindent \textbf{Scenario 5 (Varying $m_j$ and $\alpha_j$).} In this scenario we revisit the setting of independent studies. The data are generated according to Scenario 1 with $\sigma_j=1, m=1000,d=10$ and $\alpha=0.03$, but we vary $(m_j,\alpha_j)$ for the $d$ studies. To vary $m_j$, we set $m_{(1)}=\max\{m_1,\ldots,m_d\}=900$ and consider the ratio $\eta=\min\{m_1,\ldots,m_d\}/m_{(1)}$. For a given choice of $\eta$, we first sample $m_1,\ldots,m_d$ uniformly from $[\lceil m_{(1)}\eta\rceil,m_{(1)}]$ with replacement and then for each $j$, $m_j$ hypotheses are chosen at random from the $m$ hypotheses without replacement. We set $\alpha_j\in\{0.05,0.03,0.01\}$ according to $m_j\le 600$, $m_j\in(600,800]$ or $m_j>800$, respectively. Thus, in this setting studies with a higher $m_j$ have a smaller $\alpha_j$ and hence a larger weight $w_j$ on their rejections. 
Figure \ref{fig:scenario_5} reports the average FDP and the ETP for various methods as $\eta$ varies over $[0.1,0.5]$. We find that both \texttt{IRT H} and \texttt{IRT$^*$} exhibit higher power as $\eta$ increases and dominate \texttt{IRT} in power for all values of $\eta$. Furthermore, \texttt{IRT$^*$} is more powerful than \texttt{Fisher + Imputed}. When $\eta$ is large, studies receive a relatively higher weight $w_j$ on their rejections, which leads to an improved power in this setting.
\begin{figure}[!h]
\centering
\subfigure[]{
\includegraphics[width=0.8\textwidth]{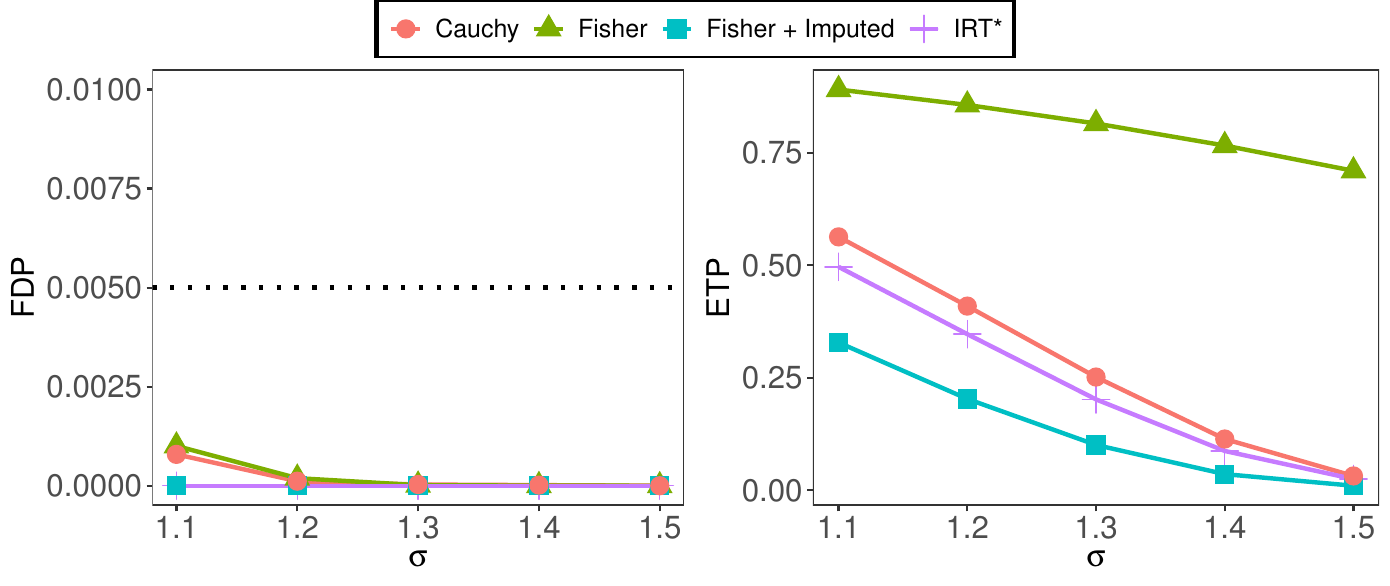}
\label{fig:scenario_8_new}}
\subfigure[]{
\includegraphics[width=0.8\textwidth]{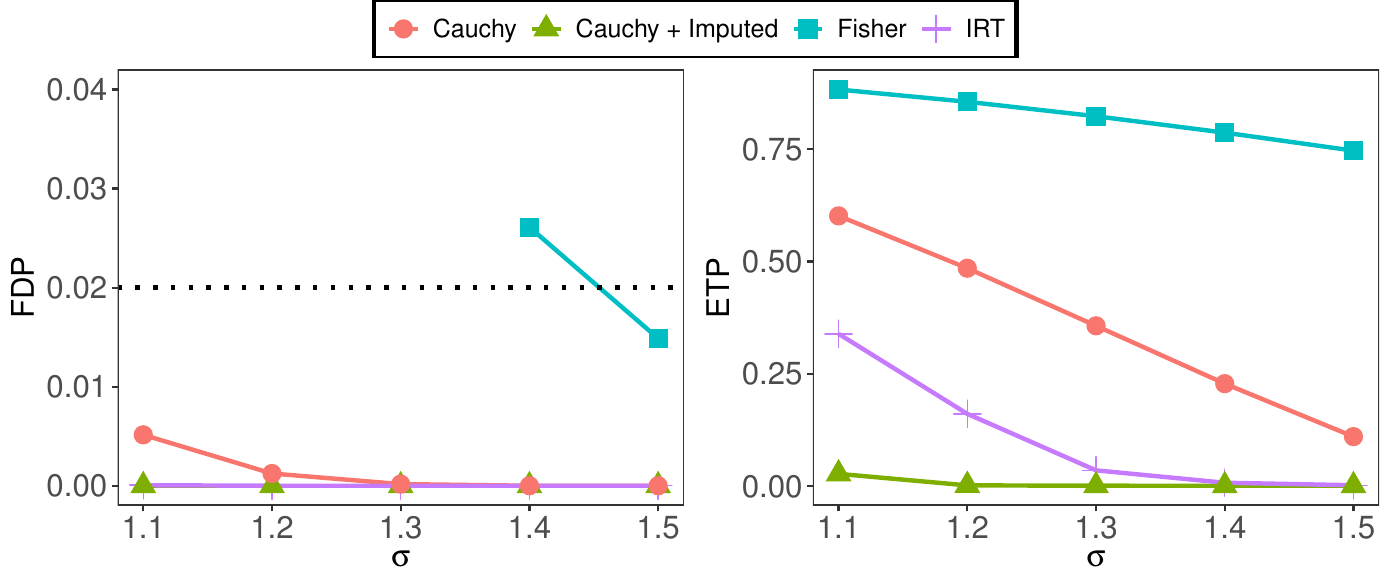}
\label{fig:scenario_7_new}} 
\caption{FDP and ETP comparisons for Scenario 6.}
\end{figure}
\\[1ex]
\noindent \textbf{Scenario 6 (conservative $p-$values: I).} We consider two settings where the study-specific $p-$values are conservative. For setting 1 we let $m_j=m=1000,\alpha_j=0.01,\alpha=0.005$ and $d=5$. For agent $j$, the summary statistics $X_{ij}\stackrel{ind.}{\sim}N(\mu_i,1)$ where $\mu_i\stackrel{i.i.d}{\sim}0.8\delta_{(0)}+0.2 N(3,1)$. To test $H_{0i}:\mu_i=0~vs~H_{1i}:\mu_i> 0$, the $p-$values are calculated using $p_{ij}=\Phi(X_{ij}/\sigma)$ where $\sigma\in\{1.1,1.2,1.3,1.4,1.5\}$. So for larger $\sigma$, the $p-$values are relatively more conservative. Figure \ref{fig:scenario_8_new} reports the average FDP and the ETP for various methods as $\sigma$ varies. We find that the power of all methods decrease as $\sigma$ increases and while \texttt{Fisher} is the most powerful across all values of $\sigma$, \texttt{IRT$^*$} and \texttt{Cauchy} exhibit similar power profiles even though the latter relies directly on the $p-$values.
			
Setting 2 borrows the design from Setting 1 but allows the studies to be correlated, i.e, $\text{Corr}(X_{ij},X_{ik})=0.5$ for all $j\ne k$, and sets $\alpha=0.02$. Since the conditions of Theorem \ref{thm3} do not hold in this setting and the corresponding $p-$values are not independent, we exclude \texttt{IRT$^*$} from our comparisons. Figure \ref{fig:scenario_7_new} reports the results of this setting and reveals that \texttt{Fisher} does not control the FDR at level $\alpha$ for all but the largest value of $\sigma$. Furthermore, \texttt{IRT} is substantially more powerful than \texttt{Cauchy + Imputed} when $\sigma$ is small but \texttt{Cauchy} dominates these two procedures in power across all values of $\sigma$.
\begin{figure}[!h]
\centering
\subfigure[]{
\includegraphics[width=0.8\textwidth]{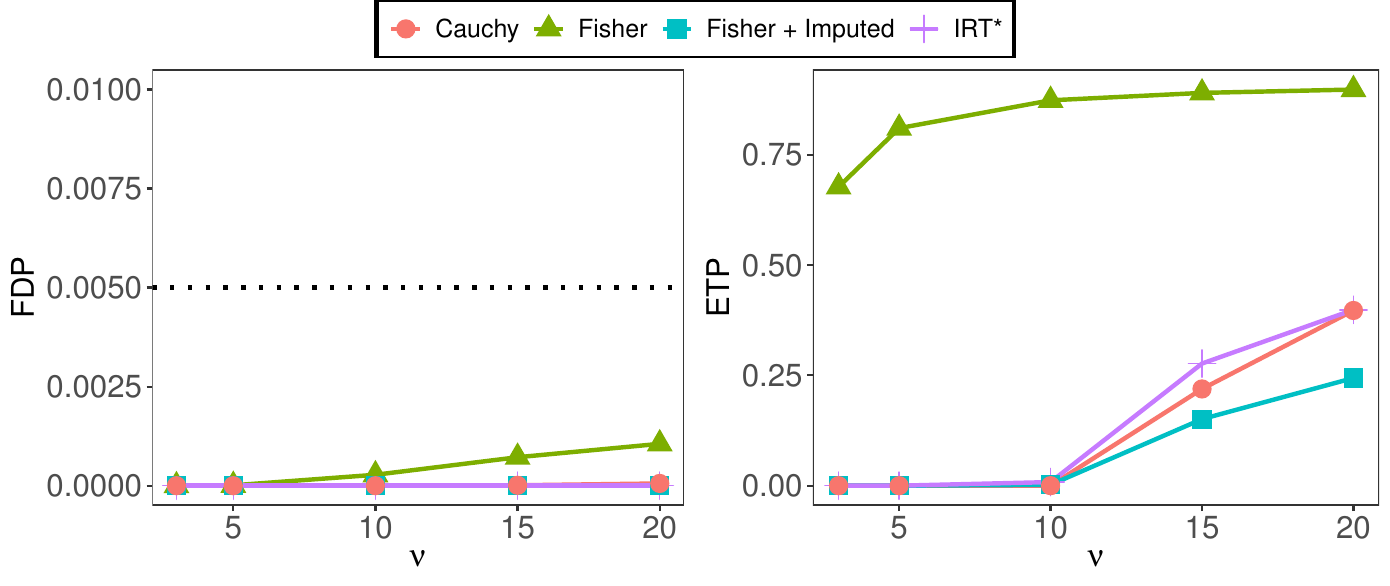}
    \label{fig:scenario_8_1_new}}
\subfigure[]{
\includegraphics[width=0.8\textwidth]{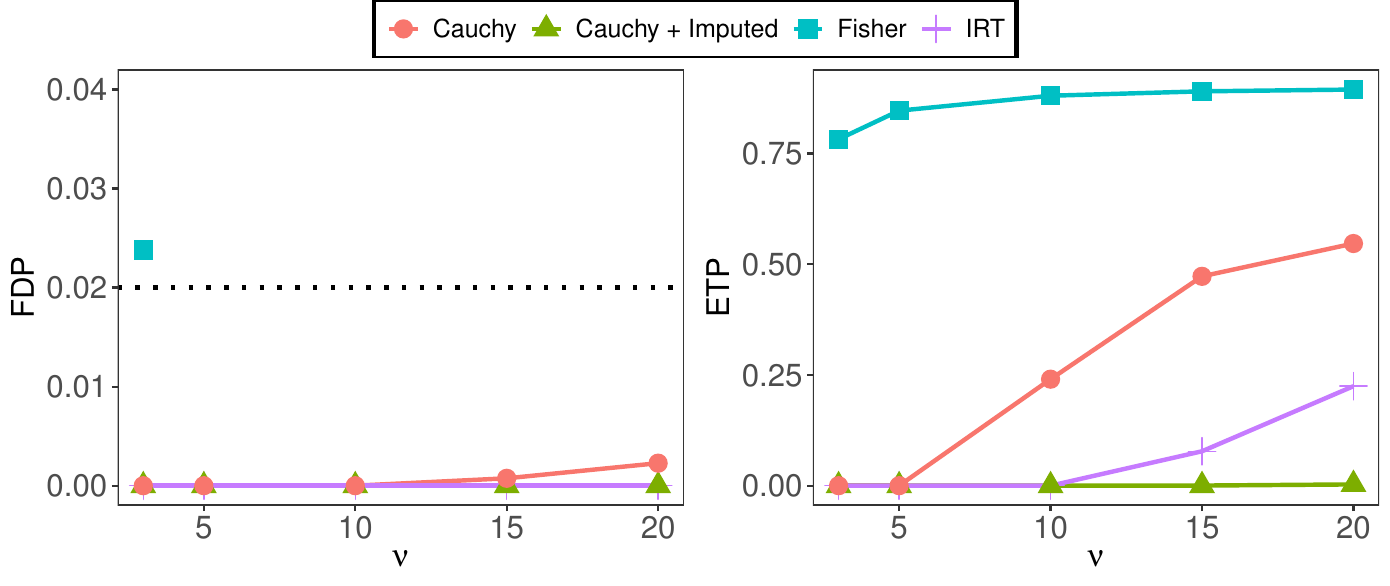}
\label{fig:scenario_7_1_new}
} 
\caption{FDP and ETP comparisons for Scenario 7.}
\end{figure}
\\[1ex]
\noindent \textbf{Scenario 7 (conservative $p-$values: II).} Here we consider two additional settings where the $p-$values are conservative. For setting 1, we borrow the design from setting 1 of Scenario 6 but compute $p_{ij}=1-F_t(X_{ij};\nu)$ where $F_t(\cdot;\nu)$ is the CDF of a central $t-$distributed random variable with $\nu$ degrees of freedom. Figure \ref{fig:scenario_8_1_new} reports the average FDP and the ETP for various methods as $\nu$ varies over $\{3,5,10,15,20\}$. All methods exhibit improved power as $\nu$ increases and \texttt{IRT*} and \texttt{Cauchy} demonstrate similar power profiles. 
In setting 2, we allow the studies to be correlated, i.e, $\text{Corr}(X_{ij},X_{ik})=0.5$ for all $j\ne k$, and set $\alpha=0.02$. We continue to exclude \texttt{IRT$^*$} from our comparisons in this setting. Figure \ref{fig:scenario_7_1_new} reports the results of this setting and reveals that \texttt{IRT} exhibits better power than \texttt{Cauchy + Imputed} when $\nu>10$. \texttt{Fisher}, in contrast, does not control the FDR for any value of $\nu$.

\begin{remark}
Note that IRT based method can sometimes have very low FDP and moderate power (for example, \textbf{Scenario 1}). 
This behavior is not an artifact but a fundamental feature of aggregating evidence from independent sources. When $\alpha<\min{\alpha_j}$,
for a true null hypothesis to be falsely rejected, it must be rejected by multiple independent studies simultaneously, an event with a much lower probability than a single false rejection. This dramatically lowers the effective error rate for false discoveries, driving the FDP to near-zero levels.
Crucially, statistical power is maintained when the underlying signals are strong enough to be detected by several studies independently. This leads to a significant overlap in the sets of true discoveries, allowing many genuine signals to pass the strict joint-rejection criterion.

This theoretical explanation is empirically validated by the results in this section. In \textbf{Scenario 1}, where studies are independent and signals are strong, all IRT variants exhibit low FDPs while \texttt{IRT*} and \texttt{IRT H} maintain high power. Conversely, in \textbf{Scenario 2}, as inter-study correlation increases, the FDP correctly begins to rise. This confirms that the near-zero FDP is a direct consequence of the independence structure, not a universal property of the method.

\end{remark}

\subsection{Performances of \texttt{IRT*} and \texttt{IRT H} when the assumptions of Lemma \ref{lem4} and Theorem \ref{thm3} are violated}
\label{sec:sims_ablation}
\begin{figure}[!h]
\centering
\includegraphics[width=0.8\textwidth]{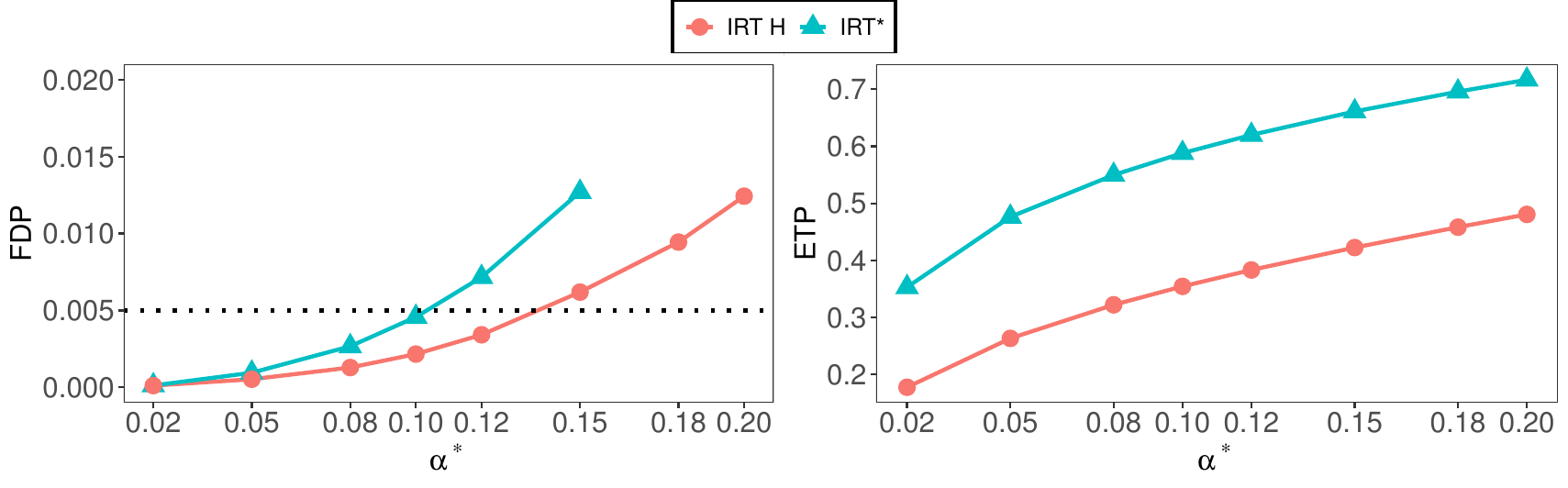}
\caption{FDP and ETP comparison for Scenario 1.}
\label{fig:ablation_1}
\end{figure}
We assess the numerical performances of \texttt{IRT*} and \texttt{IRT H} when the assumptions underlying Lemma \ref{lem4} and Theorem \ref{thm3} are violated. Specifically, we consider three scenarios. In scenario 1 the inferences from individual studies do not control the FDR at level $\alpha_j$, thus violating assumption (iii) of Lemma \ref{lem4}. In scenarios 2 and 3 the $p-$values from study $j$ are not exchangeable, which violates assumption (i) of Lemma \ref{lem4}, and the $p-$values for the $i^{th}$ testing problem are dependent, thus violating assumption (ii) of Theorem \ref{thm3}.
\\~\\
\noindent\textbf{Scenario 1 - }we borrow the independent setting from Scenario 1 of Section \ref{sec:more_sims} with $d=5$ and $\alpha=0.005$. All studies control FDR at level $\alpha^*$ but report the triplets $\{\bm\delta_j,0.01,\mathcal M\}$ to \texttt{IRT}. Thus, whenever $\alpha^*> 0.01$, the evidence indices $\bm e_j$ are no longer compound $e-$values under $\mathcal H_{0j}$. Figure \ref{fig:ablation_1} reports the average FDP and ETP across $2000$ Monte-Carlo repetitions as $\alpha^*$ varies. For large values of $\alpha^*$, both \texttt{IRT*} and \texttt{IRT H} fail to control the FDR at $0.5\%$. However when $\alpha^*$ is small, they are relatively robust to the misspecification of $\alpha_j$ as far as FDR control is concerned.
\begin{figure}[!h]
\centering
\includegraphics[width=1\textwidth]{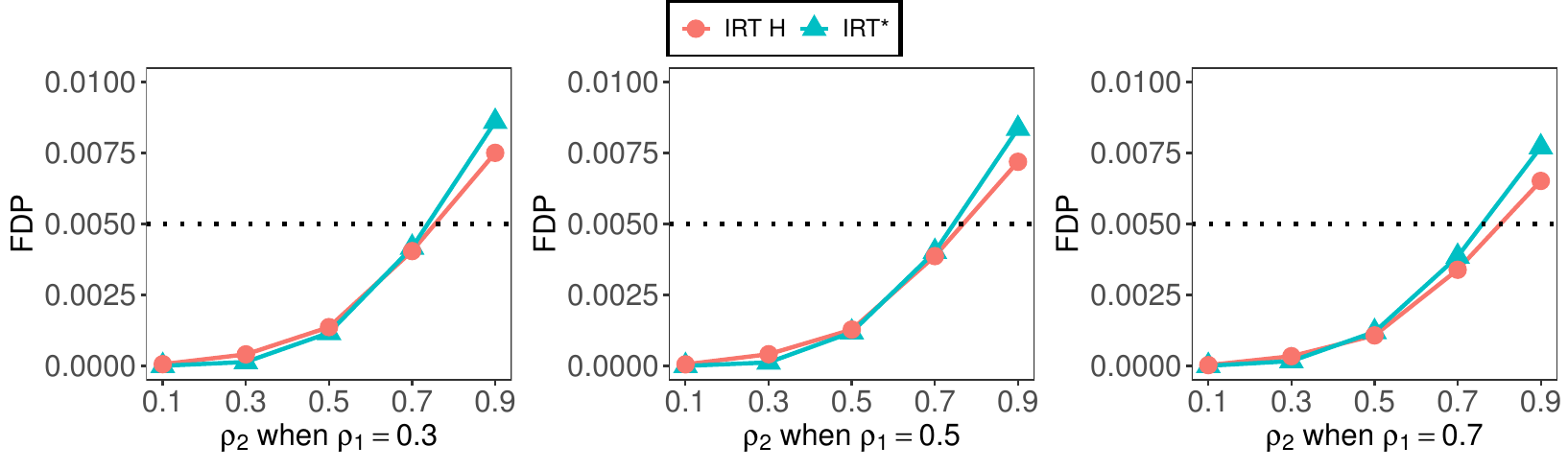}
\caption{FDP and ETP comparison for Scenario 2.}
\label{fig:ablation_2}
\end{figure}
\\[1ex]
\noindent\textbf{Scenario 2 - }we generate data according to the setting of Scenario 4 in Section \ref{sec:more_sims}. We set $d=10,~\alpha_j=0.01,~\alpha=0.005$ and introduce correlation across the studies as well as the test statistics. In particular, we let $\text{Corr}(X_{ij},X_{ik})=\rho_2,~j\ne k$ so that the $d$ $p-$values for each hypothesis are not independent unless $\rho_2 = 0$, thus violating assumption (ii) of Theorem \ref{thm3}. Furthermore, for $i\ne r\in[m]$, we set 
$$\text{Corr}(X_{ij},X_{rj})=\begin{cases}
    0,~~\text{if~} (i,r)\in\{1,\ldots,\lceil m/3\rceil\}\\
    \rho_1,~~\text{if~} (i,r)\in\{\lceil m/3\rceil+1,\ldots,2\lceil m/3\rceil\}\\
    0.9,~~\text{if~} (i,r)\in\{2\lceil m/3\rceil+1,\ldots,m\}  
\end{cases},$$ where $\lceil x\rceil$ is the smallest integer greater than or equal to $x$. Thus, the $m$ $p-$values from study $j$ are not exchangeable, which violates assumption (i) of Lemma \ref{lem4}. Figure \ref{fig:ablation_2} reports the average FDP and ETP as $\rho_2$ varies. We find that when $\rho_2$ is relatively large, both \texttt{IRT*} and \texttt{IRT H} fail to control the FDR at $0.5\%$. However, for small values of $\rho_2$, they are robust to violations of the aforementioned assumptions. Furthermore, both these methods are relatively robust to the exchangeability assumption of Lemma \ref{lem4} since increasing $\rho_1$ does not appear to have any material impact on their FDR control.
\begin{figure}[!h]
\centering
\includegraphics[width=0.8\textwidth]{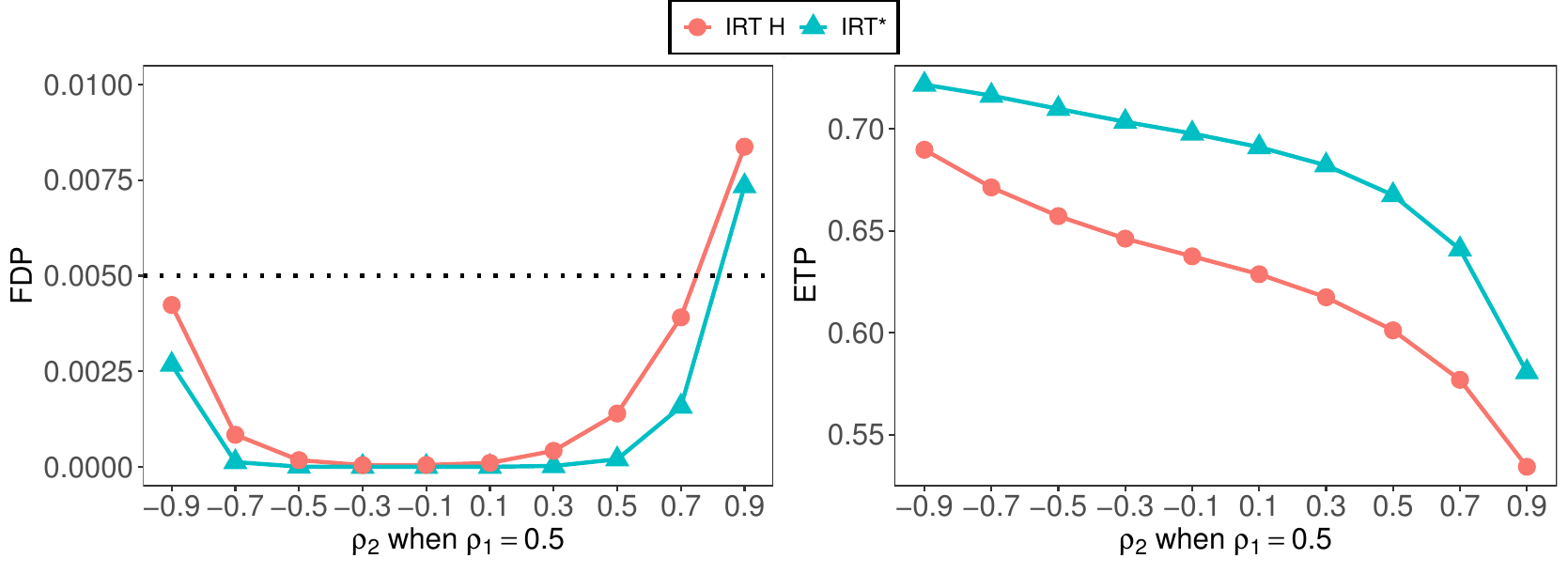}
\caption{FDP and ETP comparison for Scenario 3.}
\label{fig:ablation_3}
\end{figure}
\\[1ex]
\noindent\textbf{Scenario 3 - }we set $d=10,~\alpha_j=0.01,~\alpha=0.005$ and continue to introduce correlation across the studies as well as the test statistics. In particular, we let $\text{Corr}(X_{ij},X_{ik})=\rho_2^{|j-k|},~j\ne k$ and $\text{Corr}(X_{ij},X_{rj})=\rho_1^{|i-r|},~i\ne r$, thus imposing an AR(1) structure between the $d$ test statistics for each hypothesis and between the $m$ test statistics for each study. Figure \ref{fig:ablation_3} reports the average FDP and ETP as $\rho_2$ varies with $\rho_1=0.5$. Under this dependence structure, we find that both \texttt{IRT$^*$} and \texttt{IRT H} continue to guarantee FDR control at $\alpha$ when $\rho_2<0$, thus demonstrating robustness to violations of  assumption (ii) of Theorem \ref{thm3}. However, they fail to do so when the $d$ test statistics for each hypothesis exhibit almost perfect positive dependence, which is the case when $\rho_2=0.9$.
\section{Real data illustration}
\label{sec:realdata}
\begin{figure}[t]
				\centering
				\includegraphics[width=0.35\linewidth]{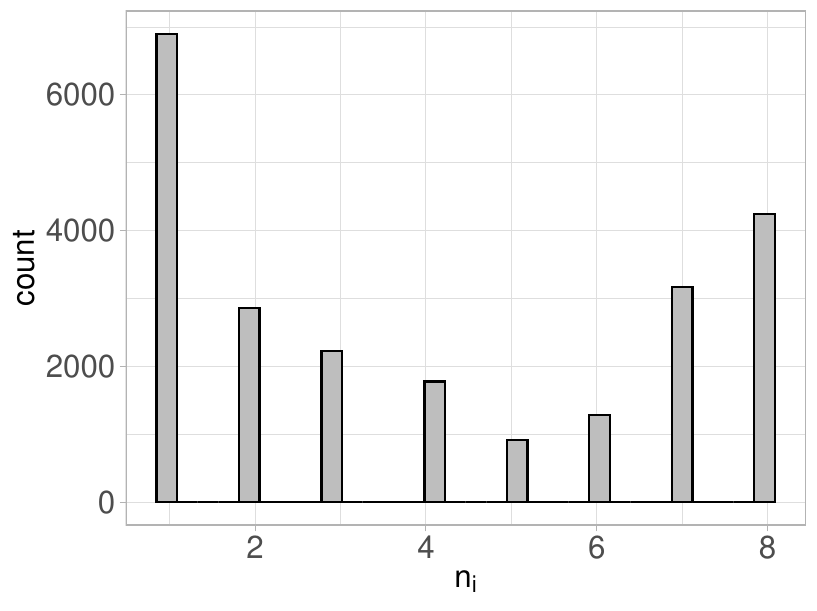}				\includegraphics[width=0.33\linewidth]{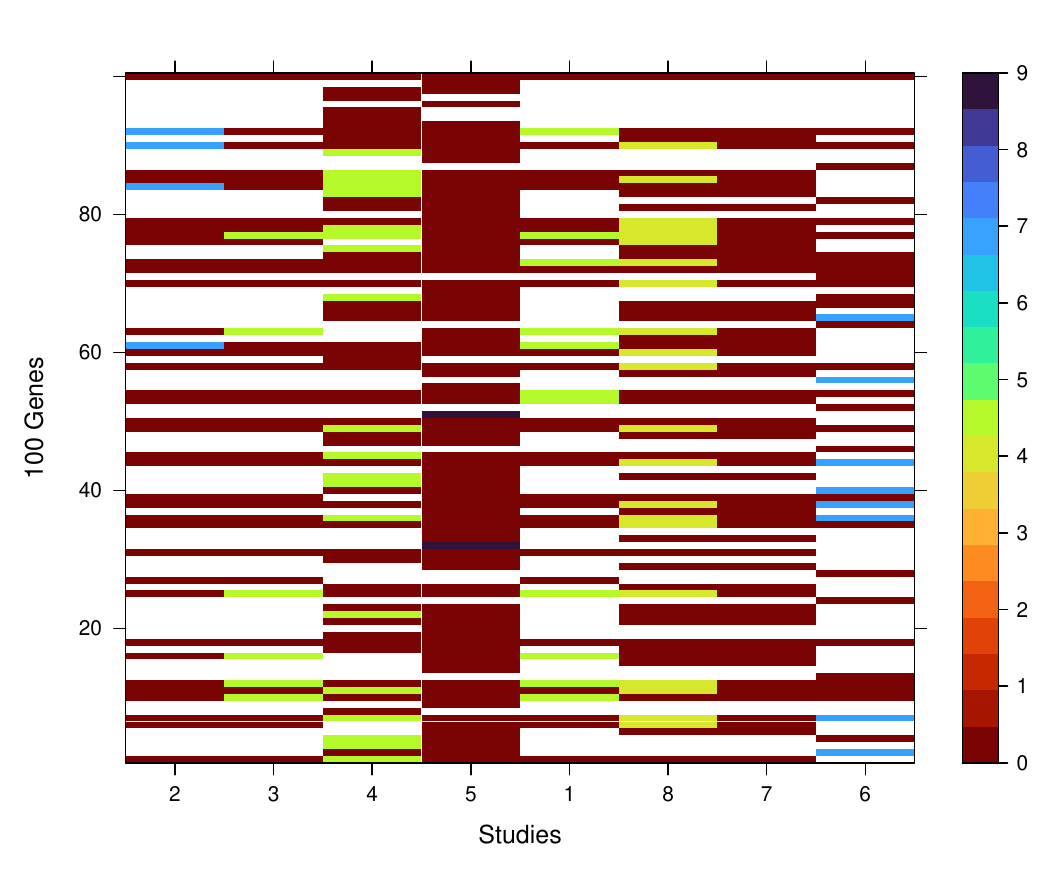}
				\caption{Left: frequency distribution of $n_i$'s. Right: heatmap of the $\log$ evidence indices for $100$ randomly sampled genes across the $d$ studies. White indicates genes not analyzed, while shades of brown represent evidence indices of $0$, indicating failure to reject.}
				\label{fig:realdata1}
			\end{figure}

We illustrate the \texttt{IRT} framework for the integrative analysis of $d=8$ microarray studies \citep{singh2002gene,welsh2001analysis,yu2004gene,lapointe2004gene,varambally2005integrative,tomlins2005recurrent,nanni2002signaling,wallace2008tumor} on the genomic profiling of human prostate cancer. The first three columns of Table \ref{tab:realdata} summarize the $d$ datasets where a total of
\begin{table}[!h]
\caption{Summary of the $d=8$ studies and the evidence against each rejected null hypothesis. Here $e_j^{+}=\max\{e_{ij}:i=1,\ldots,m_j\}$.}
\label{tab:realdata}
\centering
\scalebox{0.9}{
\begin{tabular}{ccccccc}
\toprule
$j$ & Study  & $m_j$ & Sample size & $\alpha_j$ & $\|\bm \delta_j\|_0$ & $e_{j}^{+}$ \\
\midrule
1 &\cite{singh2002gene}     &  8,799 & 102            &    0.05    &   2,094  & 84.04\\
2&\cite{welsh2001analysis}     &  8,798 & 34 & 0.01 & 921& 955.27\\
3&\cite{yu2004gene}     &  8,799 &    146         &  0.05      &   1,624 & 108.36 \\
4&\cite{lapointe2004gene}     &  13,579 &   103          &    0.05    &  3,328  & 81.60 \\
5&	\cite{varambally2005integrative}     & 19,738  &     13        &    0.01    &  282   & 6999.29\\
6&	\cite{tomlins2005recurrent}    & 9,703  &   57          &    0.01    &   1,234  & 786.30\\
7&	 \cite{nanni2002signaling}     & 12,688  &    30         &   0.01     &   0  & 0\\
8&	\cite{wallace2008tumor}     &  12,689 &  89           &   0.05     & 4,716 & 53.81\\
\bottomrule
\end{tabular}}
\end{table}
$m=23,367$ unique genes are analyzed with each gene $i$ being profiled by $n_i\in[d]$ studies. The left panel of Figure \ref{fig:realdata1} presents a frequency distribution of the $n_i$'s where almost $30\%$ of the $m$ genes are analyzed by just one of the $d$ studies while approximately $18\%$ of the genes are profiled by all $d$ studies. 
			
Our goal in this application is to use the \texttt{IRT} framework to construct a rank ordering of the $m$ gene expression profiles for prostate cancer. Such rank ordering is particularly useful when data privacy concerns prevent the sharing of study-specific summary statistics, such as $p-$values, and information regarding the operational characteristics of the multiple testing methodologies used in each study. For study $j$, our data are an $m_j\times s_j$ matrix of expression values where $s_j$ denotes the sample size in study $j$. Each sample either belongs to the control group or the treatment group and the goal is to test whether gene $i$ is differentially expressed across the two groups. Since \texttt{IRT} operates on the binary decision vector $\bm \delta_j$, we convert the expression matrices from each study to $\bm \delta_j$ as follows. For each study $j$, we first use the R-package \texttt{limma} \citep{ritchie2015limma} to get the $m_j$ vector of raw $p-$values. Thereafter, the BH procedure is applied to these raw $p-$values at FDR level $\alpha_j$ (see column five in Table \ref{tab:realdata}) to derive the final decision sequence $\bm \delta_j$. We note that typically an important intermediate step before computing the $p-$values in each study is to first validate the quality and compatibility of these studies via objective measures of quality assessment, such as \cite{kang2012metaqc}. In this application, however, we do not consider such details. 
			
			The sixth column of Table \ref{tab:realdata} reports the number of rejections for each of these studies and the last column presents the evidence against each rejected null hypothesis in study $j$. It is interesting to see that study 5 \citep{varambally2005integrative} receives the highest evidence for its rejected hypotheses, which is not surprising given the large weight $w_5$ that each of its relatively small number of rejections receives. 
			In contrast, study 8 \citep{wallace2008tumor} has the smallest non-zero evidence which is driven by the largest number of rejections reported in this study. {The right panel of Figure \ref{fig:realdata1} presents a heatmap of the $\log$ evidence indices for $100$ randomly sampled genes across the $d$ studies. Here the white shade represents a gene not analyzed by the study while the shade of brown represents an evidence index of $0$ which corresponds to a failure to reject the underlying null hypothesis. The heterogeneity across the $d$ studies is evident through the different magnitudes of the evidence indices constructed for each study.} 
			\begin{table}[t]
				\caption{Distribution of rejection overlaps across $7$ studies.}
				\label{tab:realdata3}
				\centering
				\scalebox{0.9}{
					\begin{tabular}{lccccccccccc}
						\toprule
						& &\multicolumn{7}{c}{Rejection overlap}\\ 
						\cmidrule(r){3-9}
						$j$ &  $\|\bm \delta_j\|_0$    & 1 & 2 & 3 & 4 & 5 & 6 &  8\\
						\midrule
						1 & 2,094 & -  &  509 & 1,531  & 387  &  7 &  130 &   1,029 \\
						2 & 921 &509 &  - & 423  &  108 & 1  & 27  &  324  \\
						3 &    1,624  & 1,531  & 423  &  - &  294 &  7 &  105 &  809 \\
						4 &    3,328  & 387  & 108  &  294 & -  & 17  & 172  &   970  \\
						5 &   282   &  7 &  1 &  7 & 17  &  - & 4  &  8  \\
						6 &   1,234   & 130  &  27 & 105  & 172  &  4 & -  &  365   \\
						8 &    4,716  & 1,029  &324  & 809  &  970 & 8  & 365  &   -   \\
						\bottomrule
				\end{tabular}}
			\end{table}
			Table \ref{tab:realdata3} presents the distribution of rejection overlaps 
			across the $d$ studies, with the exception of study 7. For instance, studies 1 and 3 share $1,531$ rejected hypotheses while studies 2 and 5 share just $1$ rejected hypothesis. Also, study 5, which investigates the largest number of genes, has minimal overlap with the other studies as far as its discoveries are concerned.

Since in this example study-specific $p-$values, denoted by $\{p_{ij}\}_{j\in\mathcal N_i,i\in\cM}$, are available, one can aggregate the $p-$values pertaining to each hypothesis $i$ and then determine an appropriate threshold for FDR control at level $\alpha$ using the aggregated $p-$values. However, if the underlying model is misspecified the validity of the corresponding $p-$values may be affected. In contrast, $e-$values are relatively more robust to such model misspecification \citep{wang2022false} and particularly to dependence between the $p-$values \citep{vovk2021values}. So we transform the $p-$values to $e-$values using the following calibrator from Equation (B.1) in \cite{vovk2021values}:
\[
e_{ij}^\texttt{P2E}(\kappa) = \begin{cases}
					\infty~&\text{if~}p_{ij} = 0\\
					\dfrac{\kappa(1+\kappa)^\kappa}{p_{ij}(-\log p_{ij})^{1+\kappa}}~&\text{if~}p_{ij}\in (0,e^{-\kappa-1}]\\
					0~~&\text{if~}p_{ij}\in (e^{-\kappa-1},1]
\end{cases},
\]
where we choose $\kappa=1$ following the recommendation, and write $e_{ij}^\texttt{P2E}:=e_{ij}^\texttt{P2E}(1)$. 
			{Note that $e_{ij}^\texttt{P2E}$ as defined above are bonafide e-values. Therefore, to aggregate $e_{ij}^\texttt{P2E}$ we can simply take their average}
\[
{e}_i^\texttt{\texttt{P2E},agg}=\dfrac{1}{d}\sum_{j=1}^{d}\Bigl\{e_{ij}^\texttt{P2E}~\mathbb I(i\in\mathcal M_j)+\mathbb I(i\notin \mathcal M_j)\Bigr\}.
\]
			{Furthermore, if the $p-$values $\{p_{ij}\}_{j\in\mathcal N_i}$ are independent given $\theta_i=0$ then, in the spirit of Equation \eqref{eq:eagg_alternative}, we can aggregate $e_{ij}^\texttt{P2E}$ through multiplication as follows:}
			\begin{equation*}\label{eq:p2eagg2}
				{e}_i^\texttt{\texttt{P2E},agg*}=\frac{1}{n_i}\sum_{k=1}^{n_i}\binom{n_i}{k}^{-1}\sum_{\mathcal{S}_{ki}\in \mathcal{B}_{ki}}\Big[\prod_{j\in\mathcal S_{ki}}e_{ij}^\texttt{P2E}\Big].
			\end{equation*}
In this application, we denote the method that applies the e-BH procedure on ${e}_i^\texttt{\texttt{P2E},agg}$ and ${e}_i^\texttt{\texttt{P2E},agg*}$ as \texttt{P2E} and \texttt{P2E$^*$}, respectively, and compare them to the inferences obtained from \texttt{IRT} and \texttt{IRT*}. 
\\[1.5ex]
\noindent\textbf{Ranking and thresholding using \texttt{IRT} and \texttt{P2E} - }
we aggregate the evidence indices using Equation \eqref{eq:fede} and threshold the ordered aggregated evidences using the e-BH procedure at $\alpha=0.1$. We recall that this thresholding scheme guarantees valid FDR control under unknown and arbitrary dependence between the aggregated evidences.

\begin{figure}[!h]
				\centering
				\includegraphics[width=0.9\linewidth]{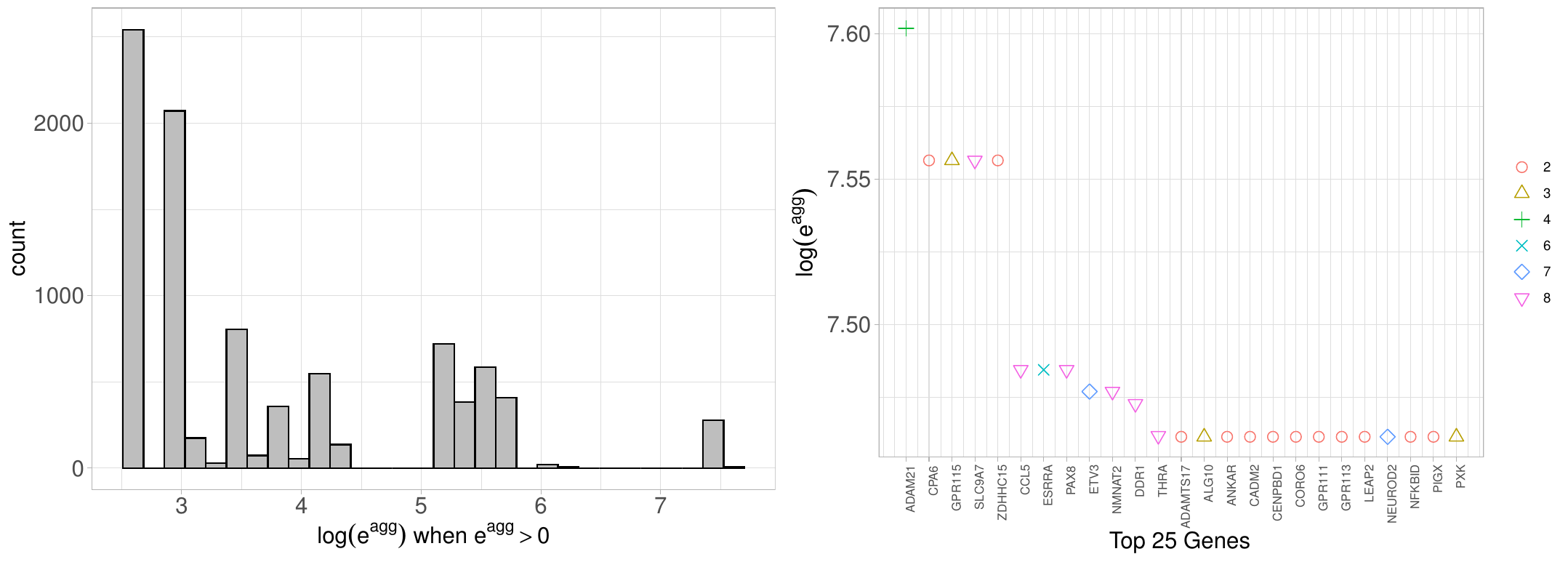}				\includegraphics[width=0.9\linewidth]{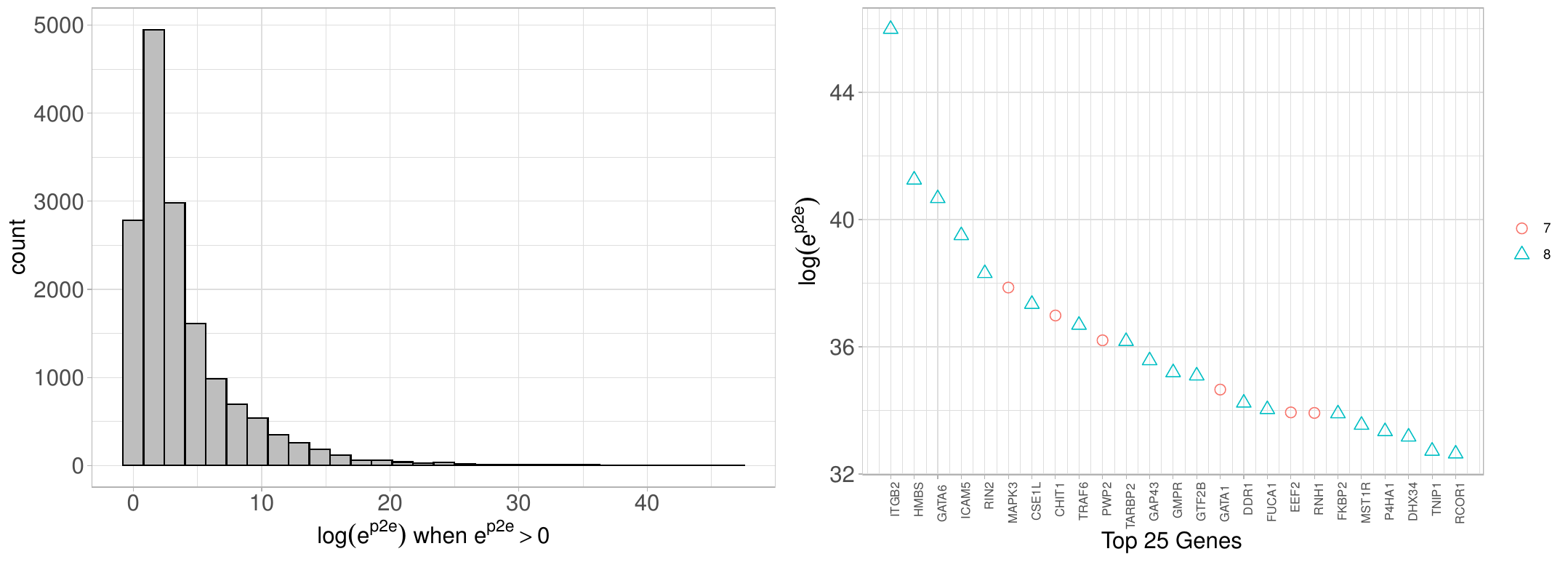}
				\caption{Left: histogram of log-transformed non-zero aggregated evidences. Right: scatter of top 25 genes, color and shape-coded by the gene analysis frequency across the $d$ studies. The top and bottom figures employ \texttt{IRT} and \texttt{P2E} respectively}
				\label{fig:realdata2}
			\end{figure}
			
			{The top-left panel of Figure \ref{fig:realdata2} presents a histogram of the log-transformed non-zero aggregated evidence from \texttt{IRT}, while the top-right panel plots the top 25 genes with respect to their aggregated evidence, colored and shape-coded by the number of times the corresponding gene was analyzed across the $d$ studies. Interestingly, the top second and third genes have $n_i=2$ and $3$, respectively, suggesting that apart from the number of times a particular null hypothesis is analyzed across the $d$ studies, the magnitude of the study-specific evidence indices also play a key role in the overall ranking. To put this into perspective, the bottom-right panel of Figure \ref{fig:realdata2} presents the top 25 genes with respect to their aggregated evidence from the \texttt{P2E} framework discussed earlier. In stark contrast to \texttt{IRT}, here the top 25 genes have $n_i\geq7$. Furthermore, both the left and right panels of Figure \ref{fig:realdata2} suggest that $e_i^\texttt{P2E}$ can be substantially larger in magnitude than $e_i^{\tt agg}$ particularly when one of the studies rejects the null hypothesis with an astronomically small $p-$value. 
				\begin{table}[!h]
					\caption{Distribution of rejected hypotheses  with respect to $n_i$ using \texttt{IRT} and \texttt{P2E} at $\alpha=0.1$.}
					\label{tab:realdata4}
					\centering
					\scalebox{0.9}{
						\begin{tabular}{cccccccccc}
							\toprule
							\multicolumn{1}{c}{}    & \multicolumn{1}{c}{\# Rejections} & \multicolumn{1}{c}{$n_i=1$} & \multicolumn{1}{c}{2} & \multicolumn{1}{c}{3} & \multicolumn{1}{c}{4} & \multicolumn{1}{c}{5} & \multicolumn{1}{c}{6} & \multicolumn{1}{c}{7} & \multicolumn{1}{c}{8} \\
							\midrule
							\multicolumn{1}{c}{\texttt{IRT}} & \multicolumn{1}{c}{2,405}              & \multicolumn{1}{c}{\bf 23.91\%}  & \multicolumn{1}{c}{2.95\%}  & \multicolumn{1}{c}{\bf 4.53\%}  & \multicolumn{1}{c}{1.25\%} & \multicolumn{1}{c}{\bf 5.03\%}  & \multicolumn{1}{c}{\bf 18.04\%}  & \multicolumn{1}{c}{15.13\%}  & \multicolumn{1}{c}{29.15\%}\\
							\midrule
							\multicolumn{1}{c}{\texttt{P2E}} & \multicolumn{1}{c}{5,336}              & \multicolumn{1}{c}{16.38\%}  & \multicolumn{1}{c}{\bf 8.24\%}  & \multicolumn{1}{c}{3.32\%}  & \multicolumn{1}{c}{\bf 3.88\%} & \multicolumn{1}{c}{2.96\%}  & \multicolumn{1}{c}{9.22\%}  & \multicolumn{1}{c}{\bf 20.48\%}  & \multicolumn{1}{c}{\bf 35.51\%}\\
							\bottomrule
					\end{tabular}}%
				\end{table}
				
				Next, we study the composition of rejected hypotheses from \texttt{IRT} and \texttt{P2E} at $\alpha=0.1$. Table \ref{tab:realdata4} presents the distribution of rejected hypotheses with respect to $n_i$ and reinforces the point that for \texttt{IRT}, the evidence weights $w_j$ play a key role in the overall ranking. 
\\[1.5ex]
\noindent\textbf{Ranking and thresholding using \texttt{IRT$^*$} and \texttt{P2E*} - }
Here we aggregate the evidence indices using the scheme discussed in Section \ref{sec:irt_star} and threshold the ordered aggregated evidences using the e-BH procedure at $\alpha=0.005$. We note that this thresholding scheme guarantees valid FDR control under (1) exchangeability of the study-specific summary statistics (Definition \ref{def:partial_exchang}), (2) symmetry of the study-specific decision rule (Definition \ref{def:symmetry}), and (3) independence of the $n_i$ summary statistics for each testing problem. In this application the summary statistics are $p-$values which are derived without any side information and so assumption (2) holds. However, verification of assumptions (1) and (3) requires additional information. Nevertheless, Figure \ref{fig:ablation_2} reveals that as far as FDR control is concerned, \texttt{IRT$^*$} is relatively robust to the violation of the exchangeability assumption (assumption (1)) and  for moderate levels of dependence between the $p-$values for each testing problem (assumption (3)), \texttt{IRT$^*$} continues to provide valid FDR control.
\begin{figure}[!h]
					\centering
\includegraphics[width=0.9\linewidth]{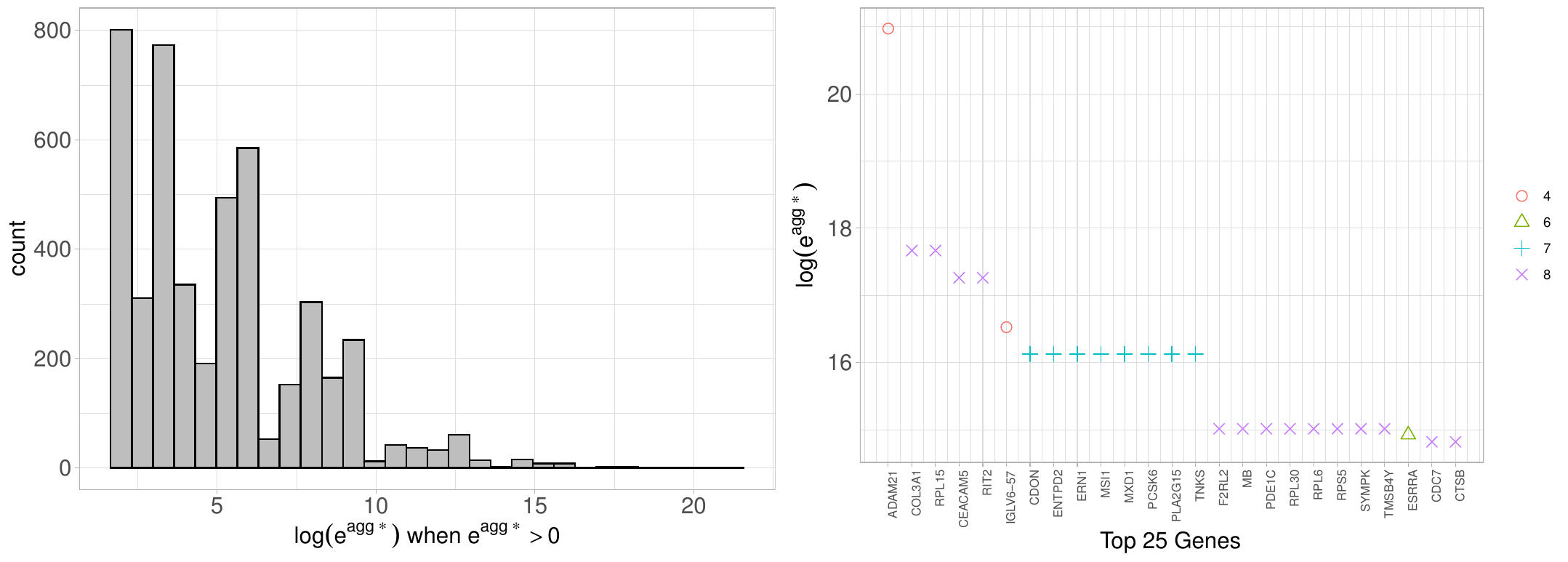}
\includegraphics[width=0.9\linewidth]{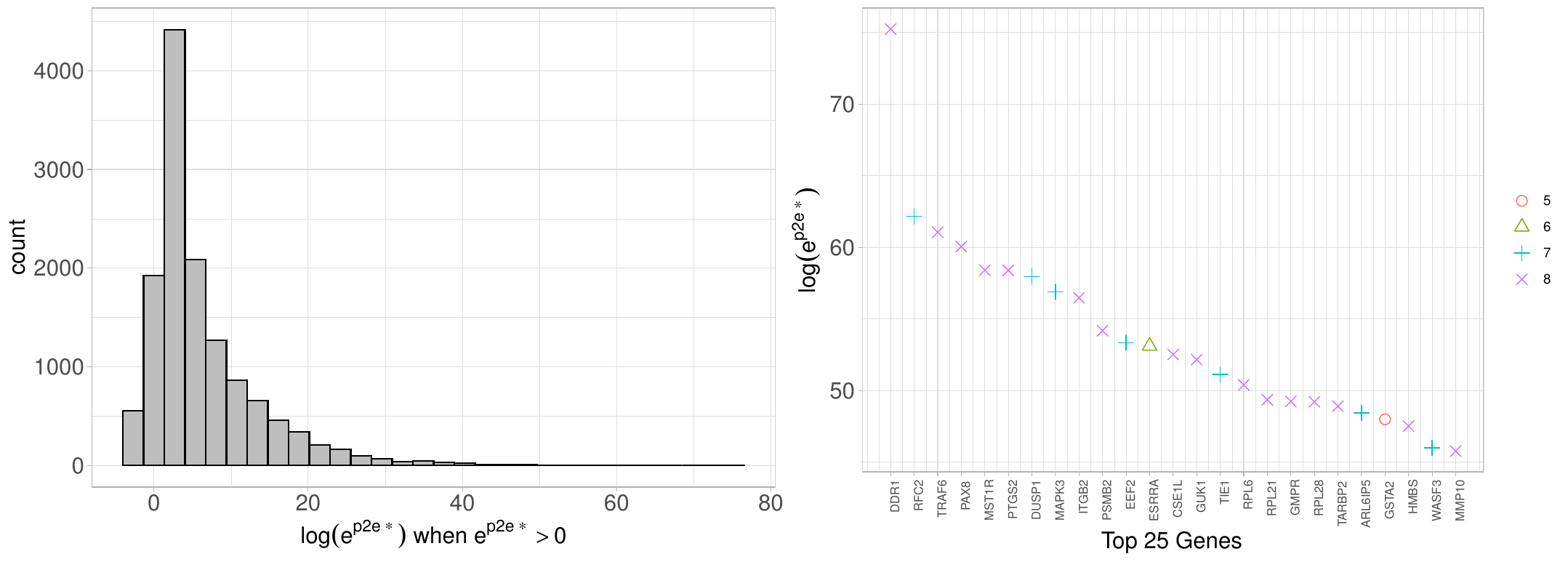}
					\caption{Left: histogram of log-transformed non-zero aggregated evidences. Right: scatter of top 25 genes, color and shape-coded by the gene analysis frequency across the $d$ studies. The top and bottom figures employ \texttt{IRT$^*$} and \texttt{P2E$^*$} respectively.}
					\label{fig:realdata_ind}
				\end{figure}
				\begin{table}[!h]
					\caption{Distribution of rejected hypotheses  with respect to $n_i$ using \texttt{IRT}$^*$ and \texttt{P2E$^*$} at $\alpha=0.005$.}
					\label{tab:realdata4_ind}
					\centering
					\scalebox{0.9}{
						\begin{tabular}{cccccccccc}
							\toprule
							\multicolumn{1}{c}{}    & \multicolumn{1}{c}{\# Rejections} & \multicolumn{1}{c}{$n_i=1$} & \multicolumn{1}{c}{2} & \multicolumn{1}{c}{3} & \multicolumn{1}{c}{4} & \multicolumn{1}{c}{5} & \multicolumn{1}{c}{6} & \multicolumn{1}{c}{7} & \multicolumn{1}{c}{8} \\
							\midrule
							\multicolumn{1}{c}{\texttt{IRT}$^*$} & \multicolumn{1}{c}{472}              & \multicolumn{1}{c}{0}  & \multicolumn{1}{c}{0}  & \multicolumn{1}{c}{\bf 2.33\%}  & \multicolumn{1}{c}{\bf 1.27\%} & \multicolumn{1}{c}{0.63\%}  & \multicolumn{1}{c}{\bf 48.52\%}  & \multicolumn{1}{c}{22.46\%}  & \multicolumn{1}{c}{24.79\%}\\
							\midrule
							\multicolumn{1}{c}{\texttt{P2E}$^*$} & \multicolumn{1}{c}{4,129}              & \multicolumn{1}{c}{\bf 4.94\%}  & \multicolumn{1}{c}{\bf 5.38\%}  & \multicolumn{1}{c}{1.98\%}  & \multicolumn{1}{c}{0.75\%} & \multicolumn{1}{c}{\bf 1.40\%}  & \multicolumn{1}{c}{12.88\%}  & \multicolumn{1}{c}{\bf 26.23\%}  & \multicolumn{1}{c}{\bf 46.43\%}\\
							\bottomrule
					\end{tabular}}%
				\end{table}
				
				The top row  of Figure \ref{fig:realdata_ind} presents a histogram of the log-transformed non-zero aggregated evidence from \texttt{IRT$^*$} (top left panel) and a plot of the top 25 genes ranked according to their aggregated evidence, colored and shape-coded by the number of times the corresponding gene was analyzed across the $d$ studies (top right panel). The bottom row presents the same plots for \texttt{P2E$^*$}. We find that \texttt{IRT}$^*$ includes three genes, ranked $1^{\tt st}, 6^{\tt th}, 23^{\tt rd}$  with $n_i\le 6$ amongst the top 25. Furthermore, \texttt{P2E$^*$} includes two genes, ranked $12^{\tt th}, 22^{\tt nd}$, with $n_i\le 6$. While this comparison is not as drastic as Figure \ref{fig:realdata2}, it continues to suggest that for \texttt{IRT}$^*$ the magnitude of study-specific evidence indices play an important role in the overall ranking. In contrast, \texttt{P2E$^*$} relies on $n_i$ and the magnitude of the $p-$values for ranking the $m$ genes. This distinction is further emphasized in Table \ref{tab:realdata4_ind} where \texttt{IRT$^*$} rejects an overall higher percentage of hypotheses than \texttt{P2E$^*$} when $n_i\le 6$.
			\end{document}